\documentclass[twocolumn,showpacs,superscriptaddress,prb,citeautoscript,amsmath,amssymb,floatfix,eqsecnum]{revtex4}
\usepackage{amsmath,amssymb}
\usepackage{bm}
\usepackage{epsfig}
\usepackage{psfrag}

\newcommand{\bd}{\bm}

\begin{document}

\title{Functional renormalization group approach to the
Anderson impurity model}

\author{Lorenz Bartosch}  
\affiliation{Institut f\"{u}r Theoretische Physik, Universit\"{a}t
  Frankfurt,  Max-von-Laue Strasse 1, 60438 Frankfurt, Germany}

\author{Hermann Freire}
\affiliation{Institut f\"{u}r Theoretische Physik, Universit\"{a}t
  Frankfurt,  Max-von-Laue Strasse 1, 60438 Frankfurt, Germany}

\author{Jose Juan Ramos Cardenas}
\affiliation{Institut f\"{u}r Theoretische Physik, Universit\"{a}t
  Frankfurt,  Max-von-Laue Strasse 1, 60438 Frankfurt, Germany}
\affiliation{Departamento Fisico-Matematico,
Universidad Autonoma de San Luis Potosi,  Ni\~{n}o Artillero 140, 
78290 San Luis Potosi, Mexico}

\author{Peter Kopietz} 
\affiliation{Institut f\"{u}r Theoretische Physik, Universit\"{a}t
  Frankfurt,  Max-von-Laue Strasse 1, 60438 Frankfurt, Germany}

\date{November 17, 2008}

 \begin{abstract}
We develop a functional renormalization group approach 
which describes the low-energy single-particle properties of
the Anderson impurity model up to intermediate on-site interactions
 $U  \lesssim 15 \Delta  $, where $\Delta$ is the hybridization in the wide-band limit. 
Our method is based on a generalization of a method
proposed by Sch\"{u}tz, Bartosch and Kopietz [Phys. Rev. B {\bf{72}}, 035107 (2005)],
using two independent
Hubbard-Stratonovich fields associated with
transverse and longitudinal spin fluctuations. Although
we do not reproduce the exponentially small Kondo scale
in the limit $U \rightarrow \infty$, the spin fluctuations included
in our approach remove the unphysical Stoner instability
predicted by mean-field theory for $U > \pi \Delta$. 
We discuss different decoupling schemes and show that a decoupling which manifestly respects the spin-rotational invariance of the problem gives rise to the lowest quasiparticle weight.
To obtain
a closed flow equation for the fermionic self-energy
we also propose a new truncation scheme of the functional renormalization group flow equations
using Dyson-Schwinger equations to express
bosonic vertex functions in terms of  fermionic  ones.

\end{abstract}


\maketitle

\section{Introduction}
The Anderson impurity model (AIM) 
is one of the most important model systems in
condensed matter physics \cite{Hewson93}. 
The model was originally proposed by Anderson \cite{Anderson61}
to describe the properties of local moments in metals. 
Its Hamiltonian describes a single
correlated impurity 
which is coupled to
a band of non-interacting conduction electrons,
 \begin{eqnarray}
 \hat{H} & = & \sum_{ \bd{k} \sigma} ( \epsilon_{ \bd{k} } - \sigma h )
 \hat{c}^{\dagger}_{ \bd{k} \sigma } \hat{c}_{ \bd{k} \sigma } 
 \nonumber  
\\
 & + & 
  \sum_{\sigma} ( E_d - \sigma h) \hat{d}^{\dagger}_{\sigma} \hat{d}_{\sigma} + U 
    \hat{d}^{\dagger}_{\uparrow} \hat{d}_{\uparrow}
  \hat{d}^{\dagger}_{\downarrow} \hat{d}_{\downarrow}
 \nonumber
 \\
 & + & \sum_{\bd{k} \sigma} ( V_{\bd{k}}^{\ast} 
 \hat{d}^{\dagger}_{ \sigma} \hat{c}_{\bd{k} \sigma } + V_{\bd{k}} 
   \hat{c}^{\dagger}_{\bd{k} \sigma } \hat{d}_{\sigma} ).
 \label{eq:Hdef}
 \end{eqnarray}
Here,
$\hat{c}_{\bd{k} \sigma}$ annihilates a non-interacting conduction electron with
momentum ${\bd{k}}$, energy dispersion $\epsilon_{\bd{k}}$
and spin projection $\sigma$,
while the operator $\hat{d}_{\sigma}$ annihilates
 a localized correlated $d$-electron with
atomic energy $E_d$ and on-site repulsion $U$.
The hybridization between 
the $d$-electrons and the conduction electrons is characterized by the
hybridization energy $V_{\bd{k}}$.
We have also included in Eq.~(\ref{eq:Hdef}) 
 the Zeemann energy $h$
associated with an external magnetic field.

The thermodynamic and spectral properties of the AIM
can be calculated accurately by means of Wilson's numerical
renormalization group \cite{Wilson75,Costi94,Hofstetter00} (NRG), see Ref.~[\onlinecite{Bulla08}]
for a recent review. However, 
the calculation of the spectral function $A ( \omega )$ of the $d$-electrons
by means of the NRG 
requires some computational effort, in particular if one needs
accurate results
for arbitrary frequencies $\omega$.
Since the quantitative  knowledge  
of the spectral function of the AIM
for all $\omega$ is essential in the context of the
dynamical mean-field theory describing
strong correlations in realistic three-dimensional fermion systems \cite{Georges96},  
it is important to develop
approximate analytical methods for calculating the spectral function
of the AIM.
Although several analytical approaches  
have been proposed  to describe the strong coupling regime
\cite{Newns83,Coleman84,MuellerHartmann84,Logan98,Hewson01,Kroha04,Janis07},
a satisfactory analytical alternative which can compete with 
the NRG in the strong coupling regime  has not been found.
This has motivated us to  develop a new
functional renormalization group (FRG) approach
to the AIM which is not based on the weak coupling
truncation used by other authors \cite{Hedden04,Karrasch08}.
Our approach extends the collective field FRG developed
in Refs.~[\onlinecite{Schuetz05,Schuetz06,Kopietz09}], which is based on the partial bosonization of
the two-body interaction using suitable
Hubbard-Stratonovich fields.
For simplicity, we focus here on
the local moment regime \cite{Hewson93}, where
the  energy $E_d$ of the $d$-level is located below the Fermi energy, but 
its double occupancy  is prohibited by a strong on-site interaction $U$.
In the limit $U \rightarrow \infty$ the low-energy properties of the AIM
are in this regime identical to those of the Kondo model describing only
the spin degree of freedom of the impurity.
It is then natural to
decouple the interaction of the AIM in terms of
Hubbard-Stratonovich fields representing collective  spin-fluctuations,
which should be treated non-perturbatively
to describe the strong-coupling regime.

Since we are interested in the correlation functions of the $d$-electrons,
we simply integrate out the conduction electrons using  the
coherent state  functional integral.
The ratio of the partition functions with and without interaction
at constant chemical potential $\mu$  and inverse temperature $\beta$ 
can then be written as
 \begin{equation}
 \frac{ \cal{Z} }{{\cal{Z}}_0} = \frac{ \int {\cal{D}} [ {d} , \bar{d} ] e^{ - S_0 [ \bar{d} , d ]  - 
 S_U [\bar{d} , {d} ] }}{  \int {\cal{D}} [ {d} , \bar{d} ] e^{ - S_0 [ \bar{d} , d ]  } }
 ,
 \label{eq:Zratio}
 \end{equation}
where the Gaussian part is given by
 \begin{eqnarray}
 S_0 [\bar{d} , d ] & = &
  \int_0^{\beta} d \tau   \sum_{\sigma} 
 \bar{d}_{\sigma} ( \tau ) \left[ 
 \partial_\tau + E_d - \mu - \sigma h \right] d_{\sigma} ( \tau ) 
 \nonumber
 \\
 & + & 
 \int_0^{\beta} d \tau \int_0^{\beta} d \tau^{\prime} \sum_{\sigma}
 \bar{d}_{\sigma} ( \tau )  \Delta^{\sigma} ( \tau - \tau^{\prime} )
  d_{\sigma} ( \tau^{\prime} ) 
 \nonumber
 \\
 &  & \hspace{-10mm} =  - \int_{ \omega} \sum_{  \sigma  }
 \left[  i \omega  - \xi^{\sigma}_0 - \Delta^{\sigma} ( i \omega ) \right]
  \bar{d}_{  \omega    \sigma}   d_{  \omega  \sigma}
 ,
 \label{eq:S0ddef}
 \end{eqnarray}
with
\begin{equation}
\xi^{\sigma}_0 = E_d - \mu - \sigma h,
 \end{equation}
and the interaction is
 \begin{equation}
 S_U [ \bar{d} , d ] = U \int_0^{\beta} d \tau \,
\bar{d}_{\uparrow} ( \tau ) d_{\uparrow} ( \tau )
 \bar{d}_{\downarrow} ( \tau ) d_{\downarrow} ( \tau ) .
 \label{eq:S1def}
\end{equation}
Here,
 $\int_{\omega} = \frac{1}{\beta} \sum_{ \omega} $
denotes  summation over fermionic Matsubara frequencies.
Later we shall take the limit $ \beta \rightarrow \infty$ where
 $\int_{\omega} =\int \frac{ d \omega}{ 2 \pi }$.
The Fourier transform of the Grassmann fields 
$d_{ \sigma} ( \tau )$ in frequency space is defined by
 \begin{equation}
 d_{\sigma} ( \tau ) = \int_{\omega}  e^{ - i \omega \tau } d_{  \omega  \sigma}
 , 
\end{equation}
and the hybridization function is
 \begin{equation}
  \Delta^{\sigma} ( \tau ) = \int_{\omega}  e^{ - i \omega \tau } \Delta^{\sigma} ( i \omega )
 , 
\end{equation}
 with Fourier components
 \begin{equation}
 \Delta^{\sigma} ( i \omega ) = 
\sum_{ \bd{k}}  \frac{ | V_{\bd{k}} |^2}{ i \omega -
 \epsilon_{\bd{k}} + \mu + \sigma h} .
 \end{equation}
The $0+1$-dimensional quantum field theory defined in Eqs.~(\ref{eq:S0ddef})--(\ref{eq:S1def})
will be the starting point of our further calculations presented below. 

Let us briefly outline the rest of this work:
In Sec.~\ref{sec:Ladder} we decouple the interaction (\ref{eq:S1def}) in the spin-singlet particle-hole channel using a complex Hubbard-Stratonovich field which describes transverse spin-flip fluctuations. 
By treating the resulting field theory on the level of the Gaussian approximation, we obtain results for the dynamic structure factor and the self-energy in the ladder approximation. Of course, only for small $U$ can we expect the ladder approximation to give reliable results. For larger couplings we find a ferromagnetic Stoner instability, consistent with the Hartree-Fock approximation but in contrast to well-established results.
To go beyond the simple ladder approximation, we use in Sec.~\ref{sec:FRG1} a collective field FRG approach for our Bose-Fermi theory and implement a simple truncation to close the set of flow equations.   
Our inclusion of transverse spin fluctuations based on the FRG removes the unphysical Stoner instability and gives results which qualitatively describe the correct physics. However, although the quasiparticle weight vanishes with increasing interaction as it should, it does not obey the well-known Kondo scaling.
To improve on these results, we introduce in Sec.~\ref{sec:twospin} a new FRG approach with partial bosonization in both the transverse and the longitudinal channel. We discuss the ambiguities with the distribution of weight among the two channels and show that in the manifestly spin-rotationally invariant case we get the strongest suppression of the quasiparticle weight. Finally, in Sec.~\ref{sec:summary} we summarize our results and discuss some open questions.

\section{Ladder approximation in the spin-singlet particle-hole channel}
\label{sec:Ladder}

\subsection{Hartree-Fock approximation and Stoner instability}
\label{subsec:HF}

If we treat the interaction (\ref{eq:S1def}) within the self-consistent 
Hartree-Fock  approximation,
we obtain the renormalized excitation energy
$\xi^{\sigma} = \xi_0^{\sigma} + \delta \xi^{\sigma}$,
with 
 \begin{equation}
  \delta \xi^{\sigma} = 
 U \int_{\omega} G_0^{{\bar\sigma}} ( i \omega ) ,
 \label{eq:xiHF}
 \end{equation}
where we have used the notation $\bar{\sigma} = - \sigma$ 
for the spin label.
The self-consistent Hartree-Fock Green function is
 \begin{equation}
 G_0^{\sigma} ( i \omega ) 
 = \frac{1}{ i \omega - \xi^{\sigma} - \Delta^{\sigma} ( i \omega ) } .
\label{eq:GHFdef}
 \end{equation}
The corresponding self-energy can be written as
 \begin{equation}
  \delta \xi^{\sigma} = 
 \frac{U}{2} [ {n} - \sigma {m} ],
 \label{eq:HFself}
 \end{equation}
where ${n} = n_{\uparrow} + n_{\downarrow}$ is the average
occupation and ${m} = n_{\uparrow} - n_{\downarrow}$
is the average magnetization of the impurity level.
For a sufficiently strong interaction,
Hartree-Fock theory predicts a finite value of $m$ even for $h \rightarrow 0$.
In the wide-band limit, where the 
hybridization function can be approximated by
 \begin{equation}
 \Delta^{\sigma} ( i \omega  ) = -i   \Delta^{\sigma}  { \rm sgn}\, \omega,
 \label{eq:wideband}
 \end{equation}
the Hartree-Fock self-consistency equation for the moment ${m}$ 
can be evaluated analytically at zero temperature. 
 For $h =0$ (where $\Delta^{\sigma} = \Delta$ is independent of the spin projection)
and in the particle-hole symmetric case
(where $E_d - \mu = -U/2$ and $n=1$) the result is 
\begin{equation}
 {m} =  \int_{\omega} \sum_{\sigma} \sigma  G_{0}^{\sigma} ( i \omega )
 = \frac{2}{\pi} \arctan \left( \frac{ U {m}}{2 \Delta} \right).
 \end{equation} 
This equation has a ferromagnetic solution ($m \neq 0$) if 
 \begin{equation}
 u_0 \equiv \frac{U}{\pi \Delta } > 1.
 \end{equation}
This ferromagnetic Stoner-instability is an artefact of the
Hartree-Fock approximation \cite{Hewson93}, which should be removed once
fluctuation corrections are properly taken into account.

\subsection{Hubbard-Stratonovich transformation in the spin-singlet particle-hole channel}
\label{subsec:HStrans}

\begin{figure}[tb]
  \centering
\epsfig{file=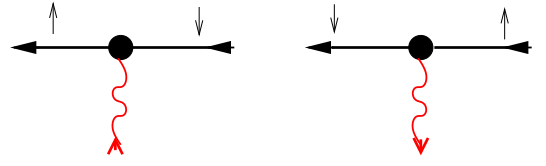,width=70mm}
\vspace{5mm}
  \vspace{-4mm}
  \caption{%
(Color online)
Graphical representation of the
bare boson-fermion vertices 
of  the 
action $S_{1}[\Phi]$  in Eq.~(\ref{eq:S1Phi}).
Solid arrows pointing into the vertices represent  $d_{\sigma}$, while outgoing
solid arrows represent $\bar{d}_{\sigma}$.
The small arrows indicate the spin projections.
The wavy arrows represent the spin-flip field $\chi$
(pointing into the black dots) and
$\bar{\chi}$ (pointing out of the black dots).
The direction of the arrows associated with the wavy lines is chosen such that
the incoming arrow adds spin  to the vertex.
}
\label{fig:barephvertex}
\end{figure}

In the strong coupling limit we expect that  
transverse spin fluctuations should play an important
role to remove the Stoner instability \cite{Logan98}.
It is therefore natural to 
decouple the interaction
in the spin-singlet particle-hole  channel with the help of a bosonic
Hubbard-Stratonovich field, describing transverse spin fluctuations. 
To this end,
we use the antisymmetry of the Grassmann fields to
write the integrand in the interaction functional  (\ref{eq:S1def}) as
 \begin{equation}
   U \bar{d}_{\uparrow}  d_{\uparrow} 
 \bar{d}_{\downarrow}  d_{\downarrow} 
 =  -  U ( \bar{d}_{\uparrow} d_{\downarrow} ) ( \bar{d}_{\downarrow} d_{\uparrow} )
 = - U \bar{s} ( \tau ) s ( \tau ),
 \label{eq:Uss}
 \end{equation}
where we have defined the composite spin-flip fields,
 \begin{equation}
 \bar{s} ( \tau )  = \bar{d}_{\uparrow} ( \tau )  d_{\downarrow} ( \tau ) \; \; ,
 \; \;  
  s ( \tau )  = \bar{d}_{\downarrow} ( \tau )  d_{\uparrow} ( \tau )  .
 \label{eq:sflipdef} 
\end{equation}
Introducing the complex bosonic Hubbard-Stratonovich fields $\chi ( \tau )$ and
$\bar{\chi} ( \tau )$ conjugate to $\bar{s} ( \tau )$ and $s ( \tau )$,
the ratio (\ref{eq:Zratio}) of the interacting and non-interacting partition functions
can be written as a functional integral over a six-component
superfield 
 $\Phi = [  
  d_{\uparrow} ,
 \bar{d}_{\uparrow} ,
 d_{ \downarrow} ,
 \bar{d}_{ \downarrow},
  \chi ,
  \bar{\chi} ]$,
\begin{equation}
 \frac{ \cal{Z} }{{\cal{Z}}_0} = \frac{ \int {\cal{D}} [ \Phi ] e^{ - S_0 [ \Phi ]  - 
 S_1 [\Phi ] }}{  \int {\cal{D}} [ \Phi ] e^{ - S_0 [ \Phi ]  } }.
 \label{eq:Zratio2}
 \end{equation}
The Gaussian part of the bare action is
 \begin{eqnarray}
 S_0 [ \Phi ] & = &
- \int_{ \omega} \sum_{  \sigma  }
 \bigl[ {G}_0^{\sigma} ( i \omega ) 
 \bigr]^{-1} \bar{d}_{ \omega \sigma}  {d}_{ \omega \sigma}  + 
 \int_{ \bar{\omega}} 
U^{-1} \bar{\chi}_{  \bar{\omega}} \chi_{  \bar{\omega}} ,
 \nonumber
 \\
 & &
 \label{eq:S0Phi}
\end{eqnarray}
and the interaction part can be written as
 \begin{eqnarray}
 S_1 [ \Phi ] & = & \int_{\bar{\omega}} \left[
\bar{s}_{ \bar{\omega}}  {\chi}_{ \bar{\omega}} + 
    s_{ \bar{\omega}} \bar{\chi}_{ \bar{\omega}}  \right] 
- \int_{\omega } \sum_{\sigma} \delta\xi^{\sigma}  
\bar{d}_{ \omega \sigma}  {d}_{ \omega \sigma}, \hspace{9mm}
\label{eq:S1Phi}
 \end{eqnarray}
where
  \begin{align}
s_{ \bar{\omega} } &= 
\int_0^{\beta} d \tau \, e^{ i \bar{\omega} \tau } s ( \tau ) = 
\int_{\omega}  
 \bar{d}_{ \omega \downarrow}
 d_{ \omega + \bar{\omega} , \uparrow}
 , \label{eq:defs} \\
\bar s_{ \bar{\omega} } &= 
\int_0^{\beta} d \tau \, e^{- i \bar{\omega} \tau } \bar s ( \tau ) = 
\int_{\omega}  
 \bar{d}_{ \omega + \bar{\omega}, \uparrow}
 d_{ \omega \downarrow} .
\label{eq:defsbar}
\end{align}
Because $[ G_0^{\sigma} ( i \omega ) ]^{-1}$ in
the Gaussian part (\ref{eq:S0Phi}) of the action is by definition
the inverse Hartree-Fock Green function (\ref{eq:GHFdef}),
the Hartree-Fock self-energy $\delta \xi^{\sigma}$ 
should be subtracted from the interaction in Eq.~(\ref{eq:S1Phi}) as a counterterm.
The boson-fermion interaction in Eq.~(\ref{eq:S1Phi}) can be written as
 \begin{eqnarray}
  \int_{\bar{\omega}} \left[
\bar{s}_{ \bar{\omega}}  {\chi}_{ \bar{\omega}} + 
 s_{ \bar{\omega}} \bar{\chi}_{ \bar{\omega}} 
 \right] 
 & = &  
 \nonumber
 \\
 & & \hspace{-33mm}
 \int_{\bar{\omega}} \int_{\omega}
 \Bigl[  
 \Gamma_0^{ (\bar{d}_{\uparrow} d_{\downarrow} \chi )} 
 (  \omega +   \bar{\omega} ,  \omega ,  \bar{\omega} )
\bar{d}_{ \omega + \bar{\omega} \uparrow} d_{ \omega  \downarrow}
 {\chi}_{ \bar{\omega}}
 \nonumber
 \\
 & & \hspace{-25mm} + 
 \Gamma_0^{ (\bar{d}_{\downarrow} d_{\uparrow} \bar{\chi} )} 
 (  \omega  - \bar{\omega},   \omega   ,  \bar{\omega} )
\bar{d}_{ \omega -  \bar{\omega}  \downarrow} d_{ \omega \uparrow}
\bar{\chi}_{ \bar{\omega}}
  \Bigr],
\label{eq:S1BF}
 \end{eqnarray}
with the bare spin-flip vertices given by
 \begin{equation}
\Gamma_0^{ (\bar{d}_{\uparrow} d_{\downarrow} \chi )}
  (  \omega +   \bar{\omega} ,  \omega ,  \bar{\omega} )
=  \Gamma_0^{ (\bar{d}_{\downarrow} d_{\uparrow} \bar{\chi} )}
 (  \omega -  \bar{\omega} ,  \omega   ,  \bar{\omega} )
 =1.
 \label{eq:barephvertex}
 \end{equation}
A graphical representation of these vertices is  shown in
Fig.~\ref{fig:barephvertex}.

\subsection{Gaussian propagator of  transverse spin-fluctuations}

\begin{figure}[tb]
  \centering
 \epsfig{file=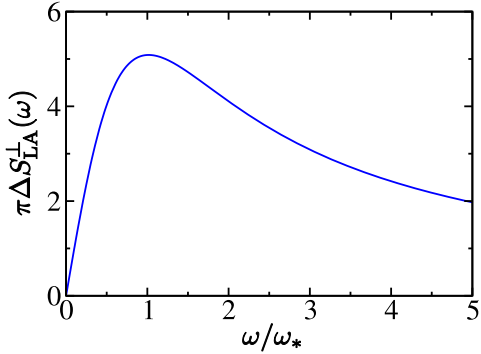,width=70mm}
  \vspace{-4mm}
  \caption{%
(Color online)
Graph of the transverse spin structure factor $S^{\bot}_{\rm LA} 
( \omega )$
in ladder approximation for 
$h=0$,
$u_0 = U / ( \pi \Delta ) = 0.9$
and particle-hole symmetry.
The characteristic energy scale of transverse spin fluctuations is
in this approximation given by
 $ \omega_\ast  \approx \Delta (1- u_0)$.
}
  \label{fig:Sbotplot}
\end{figure}

If we integrate in Eq.~(\ref{eq:Zratio2}) over the fermion fields,
we obtain the effective action 
$S_{\rm eff} [ \bar{\chi} , \chi ] $
of the spin-flip fields $\chi,\bar\chi$. 
Expanding $S_{\rm eff} [ \bar{\chi} , \chi ] $ to quadratic order in the fields
(Gaussian approximation), 
 we obtain
 \begin{eqnarray} 
 S_{\rm eff} [ \bar{\chi} , \chi ]  & \approx &
\int_{ \bar{\omega}} 
 [ F_{\rm LA}^{\rm \bot} ( i \bar{\omega} )]^{-1}
\bar{\chi}_{  \bar{\omega}} \chi_{  \bar{\omega}}.
 \label{eq:Seffchi2}
\end{eqnarray}
The inverse spin-flip propagator in 
ladder approximation  (LA) is given by
 \begin{equation}
 [ F^{\bot}_{\rm LA} ( i \bar{\omega} )]^{-1} = U^{-1}  -
 \Pi_0^{\bot} ( i \bar{\omega} ) ,
 \label{eq:FLA} 
\end{equation}
where $\Pi_0^{\bot} ( i \bar{\omega} )$ is the non-interacting 
dynamic spin-flip susceptibility
 \begin{equation}
 \Pi_0^{\bot } ( i \bar{\omega} )  = - \int_{\omega}
 G_0^{\uparrow} ( i \omega ) G_0^{\downarrow} ( i \omega - i \bar{\omega} ).
 \label{eq:pibot}
 \end{equation}
In the wide-band limit, where $\Delta^{\sigma} ( i \omega )$ is given by
Eq.~(\ref{eq:wideband}), the integration in Eq.~(\ref{eq:pibot}) can be performed 
analytically at zero temperature.
In  this work we only need $\Pi_0^{\bot } ( i \bar{\omega} ) $
in the absence of an external magnetic field and spontaneous
ferromagnetism, where
$\xi^{\sigma} = \xi$ and $\Delta^{\sigma} = \Delta$ are independent
of the spin projection $\sigma$,
 \begin{eqnarray}
  \Pi_0^{\bot} ( i \bar{\omega} )
 = \frac{\Delta} {\pi | \bar{\omega} | ( | \bar{\omega} | +
 2 \Delta ) } \ln \left[
 \frac{ \xi^2 + ( | \bar{\omega} | + \Delta )^2}{ \xi^2 + \Delta^2 } \right].
 \hspace{5mm}
 \label{eq:Pi0ph}
 \end{eqnarray}
If we assume in addition particle-hole symmetry so that
$\xi= \xi_0 + \delta \xi =0$, then
Eq.~(\ref{eq:Pi0ph}) further simplifies to 
 \begin{eqnarray}
  \pi \Delta \Pi_0^{\rm \bot} ( i \bar{\omega} )
 & = &  
\frac{ \ln [ 1 + | \bar{\epsilon} |]   }{ | \bar{\epsilon} | ( 1 + | \bar{\epsilon} |/2) }
 \nonumber
 \\
 & = & 1 - | \bar{\epsilon} | + \frac{5}{6} \bar{\epsilon}^2 + \mathcal{O} ( | \bar{\epsilon} |^3 ),
 \label{eq:Piflipsmall} 
\end{eqnarray}
where $\bar{\epsilon} = \bar{\omega} / \Delta$. Within the LA
the spectral density $S_{\rm LA}^{\bot} ( \omega )$ of transverse spin-fluctuations
(the dynamic structure factor) is then given  by 
 \begin{equation}
\Pi^{\bot }_{\rm LA} ( i \bar{\omega} ) =
 \frac{\Pi_0^{\bot } ( i \bar{\omega} )}{ 1 -  U
\Pi_0^{\bot} ( i \bar{\omega} )} =
\int_0^{\infty}  \frac{d \omega}{\pi}   
S_{\rm LA}^{ \bot} ( \omega )
 \frac{ 2 \omega}{ \omega^{ 2} + \bar{\omega}^2 },
 \label{eq:SLdef}
 \end{equation}
or, equivalently,
 \begin{eqnarray}
  {\rm Im}\, \Pi_{\rm LA}^{\bot} (  \omega + i 0 )  & = &
 {\rm sgn}\, \omega\,
 S_{\rm LA}^{\bot} ( | \omega | ).
 \end{eqnarray}
A graph of $S_{\rm LA}^{ \bot} ( \omega )$ for
$ u_0 =U /( \pi \Delta ) =0.9 $ is shown in Fig.~\ref{fig:Sbotplot}.
In the regime $ 0 < 1- u_0 \ll 1$ the LA predicts a
well defined peak in the dynamic structure factor
at the energy scale 
 \begin{equation}
\omega_{\ast} = \Delta ( 1 - u_0).
 \end{equation}
The width of the peak is of the order of $\Delta$.
In fact,  in this regime the low-frequency behavior of
$S^{\bot}_{\rm LA} 
( \omega )$ can easily be obtained analytically.
For $ | \omega | \ll \Delta$ we find
 \begin{equation}
\Pi_{\rm LA}^{\bot} ( i \bar{\omega} ) = \frac{1}{\pi}
 \frac{ 1}{ \omega_{\ast} + u_0 | \bar{\omega} | },
 \end{equation}
so that  the corresponding structure factor is
 \begin{equation}
 S_{\rm LA}^{\bot } ( \omega )  \approx \frac{1}{\pi}
 \frac{   \omega}{ \omega_{\ast}^2 + \omega^2 }.
 \end{equation}  
For $u_0 \rightarrow 1$ the energy scale $\omega_{\ast}$ vanishes, 
indicating an instability towards spontaneous ferromagnetism,
as suggested by the Hartree-Fock approximation discussed 
in Sec.~\ref{subsec:HF}.
Because we know that the AIM does not
exhibit spontaneous ferromagnetism for arbitrary 
values of $U$, we expect that the LA is only accurate 
for $u_0 \ll 1$.
Before employing more sophisticated FRG methods to include  fluctuation corrections
which remove this unphysical Stoner instability, it is instructive 
to consider the $d$-electron self-energy within the LA, which we shall do
in the following subsection.

\subsection{Fermionic self-energy in ladder approximation}

\begin{figure}[tb]
  \centering
 \epsfig{file=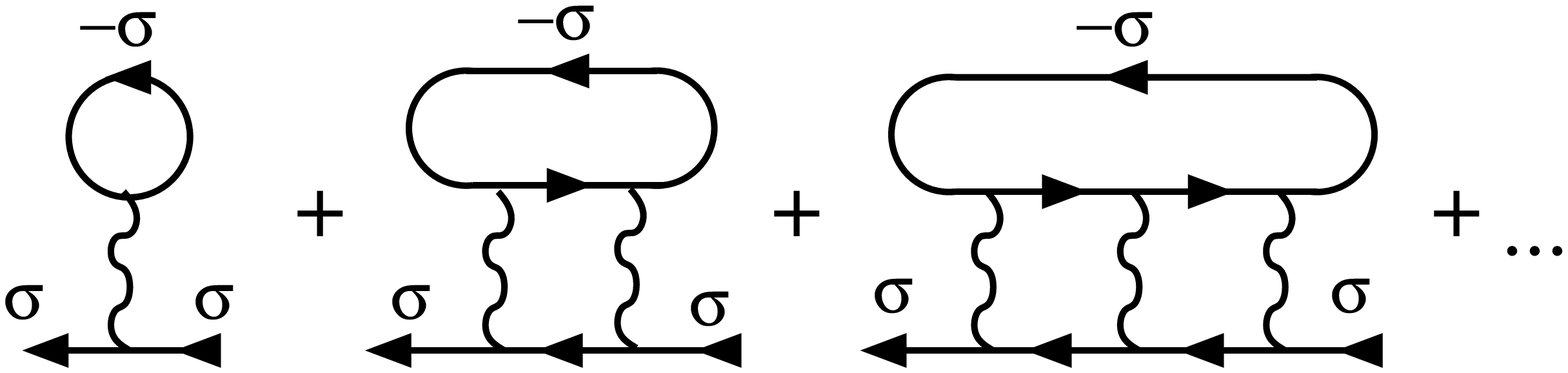,width=80mm}
  \caption{%
Diagrams contributing to  the fermionic self-energy  $\Sigma ( i \omega )$
in particle-hole ladder approximation, see Eq.~(\ref{eq:selfladder}).
Solid arrows denote fermionic Hartree-Fock Green functions 
and wavy lines denote the bare interaction.
The first-order (Hartree) diagram is subtracted in 
Eq.~(\ref{eq:sigmaLA}).
}
  \label{fig:phLA}
\end{figure}

\begin{figure}[tb]
  \centering
\epsfig{file=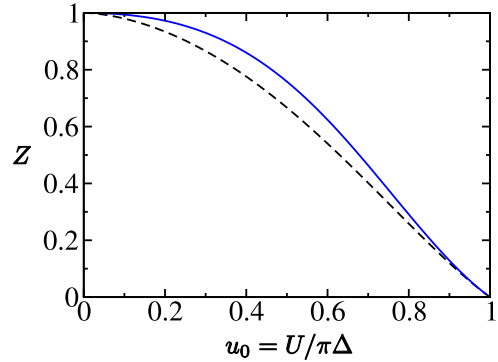,width=70mm}
  \caption{%
(Color online)
Graph of the quasiparticle residue  $ Z$ 
as a function of $u_0 = U / ( \pi \Delta )$
within LA for $h =0$ and particle-hole symmetric filling $E_d - \mu = - \frac{U}{2}$. While in the calculation of the solid line (blue) the full polarization was used, in the calculation of the dashed line (black) we have used the linearized spin susceptibility $\pi \Delta \Pi_0^{\rm \bot} ( i \bar{\omega} ) \approx 1 - | \bar{\omega} |/\Delta$.
}
  \label{fig:ZUplot}
\end{figure}

The  LA in the particle-hole channel gives the following   self-energy 
for the $d$-electrons,
 \begin{equation}
 \Sigma^{\sigma} ( i \omega ) = \int_{\bar{\omega}} F_{\rm LA}^{\bot } (
i \bar{\omega} ) G^{\bar{\sigma}}_0 ( i \omega -  i \sigma \bar{\omega} ).
 \label{eq:selfladder} 
\end{equation}
The infinite series of Feynman diagrams which is included
in  Eq.~(\ref{eq:selfladder}) is shown in Fig.~\ref{fig:phLA}.
Subtracting the Hartree-Fock correction $\delta \xi^{\sigma}$
given in Eq.~(\ref{eq:xiHF}) which formally arises from the
counterterm in Eq.~(\ref{eq:S1Phi}), the frequency-dependent part of the
self-energy can also be written as
 \begin{equation}
\delta \Sigma^{\sigma} ( i \omega )   \equiv \Sigma^{\sigma} ( i \omega ) -
\delta \xi^{\sigma} =
   U^2
 \int_{\bar{\omega}}  \Pi^{\bot}_{\rm LA} ( i \bar{\omega} )
G^{\bar{\sigma}}_0 ( i \omega -  i {\sigma} \bar{\omega} ).
 \label{eq:sigmaLA}
 \end{equation}
Of particular interest is the quasiparticle residue $Z^{\sigma}$, which is defined by
\begin{equation}
 Z^{\sigma}  = \frac{1}{ 1 -  \left.
\frac{ \partial {\rm Re}\, \Sigma^{\sigma} ( \omega + i 0 ) }{
 \partial \omega } \right|_{\omega =0} }.
 \label{eq:Zomegadef}
 \end{equation}
In Fig.~\ref{fig:ZUplot} we show the predication of the LA for the
interaction-dependence 
of  the quasiparticle residue  in the non-magnetic, particle-hole symmetric case with and without linearization of the spin-flip susceptibility $\Pi_0^{\rm \bot} ( i \bar{\omega} )$.
Obviously, the linearization of the spin-flip susceptibility does lead to modest changes of the $Z$-factor.
The vanishing of $Z$ at $u_0 =1$ is an unphysical artefact of the LA, which
implicitly contains the Stoner instability via the spin susceptibility.
In the following section we shall show how 
to remove this instability by including the feedback of the fermionic wave function
renormalization on the spin susceptibility.

\section{FRG with
partial bosonization in the spin-singlet particle-hole channel}
\label{sec:FRG1}

To go beyond the ladder approximation, 
we now use the collective field FRG approach 
developed
in Refs.~[\onlinecite{Schuetz05,Schuetz06,Kopietz09}] to study the
mixed Bose-Fermi theory defined by the action
$S [ \Phi ] = S_0 [ \Phi ] + S_1 [ \Phi ]$ in 
Eqs.~(\ref{eq:S0Phi}) and (\ref{eq:S1Phi}).
In fact, this theory has a formal similarity with the
theory describing two coupled metallic chains, 
which was studied by means of the FRG  in Ref.~[\onlinecite{Ledowski07}]. 
The exact FRG flow equations in the present problem 
are therefore represented by the same Feynman diagrams as given 
in  Ref.~[\onlinecite{Ledowski07}].

\subsection{Exact FRG flow equations}

\begin{figure}[tb]
  \centering
\epsfig{file=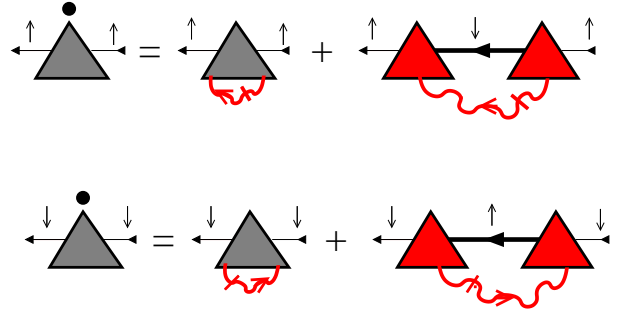,width=80mm}
\vspace{5mm}
  \vspace{-4mm}
  \caption{%
(Color online)
Exact flow equation for the fermionic self-energy $\Sigma_{\Lambda}^{\sigma}(\omega)$ 
with Hubbard-Stratonovich decoupling in the spin-singlet particle-hole channel.
In our  cutoff scheme only the  bosonic propagator is regularized via a cutoff.
Thick black arrows denote the flowing fermion propagator given in
Eq.~(\ref{eq:Gflow}), while
thick wavy arrows with a slash denote the flowing single-scale spin-flip propagator
defined in Eq.~(\ref{eq:dotFbot}). The irreducible vertices are denoted by shaded triangles 
with the appropriate number of external legs. The dots over the fermionic 
two-point vertices on the left-hand side represent the derivative with respect to the
RG cutoff $\Lambda$.
}
\label{fig:ph-selfenergy}
\end{figure}

To derive exact FRG  flow equations, 
we modify our original model
by introducing an infrared cutoff $\Lambda$ into the Gaussian part
(\ref{eq:S0Phi}) of the action.
In our renormalization group (RG) scheme 
the cutoff is  introduced 
only in the bosonic sector \cite{Schuetz05,Schuetz06,Kopietz09,Ledowski07}.
In previous FRG studies of the AIM, fermionic degrees of freedom were directly integrated out and a sharp Matsubara frequency 
cutoff was employed \cite{Hedden04,Karrasch08}. For our purpose it is better to work with
a smooth cutoff in order to avoid artificial singularities introduced by a 
sharp cutoff. Formally, we introduce  the cutoff  
via the following substitution in the bosonic part of the Gaussian action
$S_0 [ \Phi ]$ given in Eq.~(\ref{eq:S0Phi}),
 \begin{equation}
 U^{-1} \rightarrow  U^{-1} + R_{\Lambda} (  i  \bar{\omega}  ),
 \label{eq:UR} 
\end{equation}
where
 \begin{equation}
R_{\Lambda} ( i \bar{\omega}  ) = \frac{ \Lambda}{\pi \Delta^2   }  
  R ( | \bar{\omega} | / \Lambda) ,
 \label{eq:RRdef} 
\end{equation}
and the function $R (x )$ is given by\cite{Litim01}
 \begin{equation}
R (x) = (1-x) \Theta (1-x ).
 \label{eq:Litim} 
\end{equation} 
The flowing  spin-flip propagator is then
 \begin{eqnarray}
 F^{\bot}_{\Lambda} ( i \bar{\omega} ) & = &  
[ U^{-1} -  \Pi^{\bot }_{\Lambda} ( i \bar{\omega} ) +
R_{\Lambda}   ( i  \bar{\omega}  ) ]^{-1}
\nonumber
 \\
 & = &
\frac{U}{ 1 + U 
\left[ R_{\Lambda}   ( i  \bar{\omega}  )    -   \Pi^{\bot }_{\Lambda} ( i \bar{\omega} ) 
 \right]},
 \label{eq:Fbot}
 \end{eqnarray}
where $\Pi^{\bot }_{\Lambda} ( i \bar{\omega} )$ is the flowing irreducible
transverse spin susceptibility.
Introducing the corresponding single-scale propagator
 \begin{equation}
 \dot{F}^{\bot}_{\Lambda} ( i \bar{\omega} ) = [ - \partial_{\Lambda}
 R_{\Lambda} ( i \bar{\omega}  ) ] [ F^{\bot}_{\Lambda} ( i \bar{\omega} ) ]^2,
 \label{eq:dotFbot} 
\end{equation}
the exact FRG  flow equation for the irreducible self-energy of
spin-$\sigma$ electrons can then be written as
\begin{eqnarray} 
\partial_{\Lambda} \Sigma^{\sigma}_{\Lambda}(i\omega) & = &
\int_{\bar{\omega}} 
\dot{F}^{\bot}_{\Lambda} (i\bar{\omega})
\Gamma_{\Lambda}^{(\bar{d}_{\sigma} d_{\sigma} \bar{\chi} \chi)}
( \omega , \omega ; \bar{\omega},  \bar{\omega}) \hspace{15mm}
\nonumber \\
&   + & 
 \int_{\bar{\omega}} \dot{F}^{\bot}_{\Lambda} (i\bar{\omega})
G_{\Lambda}^{\bar{\sigma}} (i\omega-i \sigma \bar{\omega})
 \nonumber \\
 & & \times
\Gamma_{\Lambda}^{ (\bar{d}_{\sigma} d_{\bar{\sigma}} {\chi}_{\sigma}  )}
(\omega, \omega  - \sigma \bar{\omega}  , \sigma \bar{\omega})
\nonumber \\
& & 
 \times 
\Gamma_{\Lambda}^{  (\bar{d}_{\bar{\sigma}} d_{\sigma} {\chi}_{\bar{\sigma}} )  }
( \omega - \sigma \bar{\omega},  \omega, \sigma \bar{\omega}) .
 \label{eq:selfflow}
 \end{eqnarray}
Here, $\Gamma_{\Lambda}^{(\bar{d}_{\sigma} d_{\sigma} \bar{\chi} \chi)}
( \omega , \omega ; \bar{\omega},  \bar{\omega})$ is the flowing irreducible vertex with
two bosonic and two fermionic external legs, and
 for the labels of the flowing three-legged boson-fermion vertices 
we have used the
short notation $\chi_{\uparrow} = \chi$ and $\chi_{\downarrow}  = \bar{\chi}$, so that
the non-zero combinations are
$
\Gamma_{\Lambda}^{ (\bar{d}_{\uparrow} d_{\downarrow} {\chi}_{\uparrow}  )}
(\omega, \omega  -  \bar{\omega}  ,  \bar{\omega})
= \Gamma_{\Lambda}^{ (\bar{d}_{\uparrow} d_{\downarrow} {\chi}  )}
(\omega, \omega  -  \bar{\omega}  ,  \bar{\omega})$ and
$
\Gamma_{\Lambda}^{ (\bar{d}_{\downarrow} d_{\uparrow} {\chi}_{\downarrow}  )}
(\omega  -  \bar{\omega}  , \omega,  \bar{\omega})
= \Gamma_{\Lambda}^{ (\bar{d}_{\downarrow} d_{\uparrow} \bar{\chi}  )}
( \omega  -  \bar{\omega}  , \omega , \bar{\omega})$.
Actually, the three-legged boson-fermion vertices have the symmetry
 \begin{equation}
 \Gamma^{(\bar{d}_{\downarrow} d_{\uparrow} \bar{\chi} )}_{\Lambda} 
(  \omega -  \bar{\omega} ,  \omega ,  \bar{\omega} )  = 
 \Gamma^{(\bar{d}_{\uparrow} d_{\downarrow} {\chi} )}_{\Lambda} 
(  \omega , \omega -  \bar{\omega} ,   \bar{\omega} ) ,
 \label{eq:vertexsym}
 \end{equation}
so that the two vertices in Eq.~(\ref{eq:selfflow}) have the same value.
The flowing $d$-level Green function $G^{\sigma}_{\Lambda} (i \omega )$ is related to
the flowing self-energy via the Dyson equation,
 \begin{eqnarray}
 {G}_\Lambda^{\sigma} ( i \omega ) & = & \frac{1}{ i \omega  - {\xi}_0^{\sigma}  - 
\Delta^{\sigma} 
( i \omega )   -  \Sigma_{\Lambda}^{\sigma} ( i \omega ) }.
 \label{eq:Gflow} 
\end{eqnarray}
A graphical representation of the exact FRG flow equation
(\ref{eq:selfflow}) is shown in Fig.~\ref{fig:ph-selfenergy}. 
The initial values of the boson-fermion vertices are equal to the bare 
vertices given in Eq.~(\ref{eq:barephvertex}), which is unity with our normalization.
In principle, the integration of the FRG flow equation
 (\ref{eq:selfflow}) generates also the Hartree-Fock contribution
 (\ref{eq:HFself}) to the self-energy as a boundary term which is of first-order 
in $U$ and appears
when  we integrate this equation up to $\Lambda =0$.
In practice, it is better to drop this first-order term
and include the Hartree-Fock self-energy from the beginning
into the $d$-electron propagator, which amounts to imposing the
initial condition
 \begin{equation}
\Sigma_{\Lambda_0}^{\sigma} ( i \omega ) =\delta \xi^{\sigma} = 
 \frac{U}{2} [ {n} - \sigma {m} ],
 \label{eq:Sigmainitial} 
\end{equation}
so that the initial $G_{\Lambda_0}^{\sigma} ( i \omega )$ is the
Hartree-Fock Green function.

\subsection{Truncation of the FRG equations via  Dyson-Schwinger equations}
\label{subsec:trunc}

To obtain a closed system of RG flow equations, 
we need additional RG equations for the
four-legged boson-fermion vertex
$\Gamma_{\Lambda}^{(\bar{d}_{\sigma} d_{\sigma} \bar{\chi} \chi)}$,
for the three-legged boson-fermion vertices
 $\Gamma_{\Lambda}^{ (\bar{d}_{\uparrow} d_{\downarrow} {\chi}  )}$ and
 $\Gamma_{\Lambda}^{ (\bar{d}_{\downarrow} d_{\uparrow}  \bar{\chi}  )}$, as well as for the flowing irreducible spin-flip susceptibility 
$\Pi^{\bot}_{\Lambda} ( i \bar{\omega} )$
which determines the single-scale spin-flip propagator
$\dot{F}^{\bot}_{\Lambda} (i\bar{\omega})$.
The exact FRG flow equations for the three- and four-legged boson-fermion vertices
have been written down diagrammatically in Ref.~[\onlinecite{Ledowski07}]; the
crucial point is that the right-hand sides of these flow equations vanish at
the initial scale $\Lambda = \Lambda_0$  because they depend on higher order vertices which are
not contained in the bare action.
 It is therefore reasonable to
ignore the RG flow of the three- and four-legged boson-fermion vertices,
which amounts to truncating the flow equation (\ref{eq:selfflow}) by
replacing these vertices by
their initial values,
 \begin{subequations}
 \begin{eqnarray}
\Gamma_{\Lambda}^{(\bar{d}_{\sigma} d_{\sigma} \bar{\chi} \chi)}
( \omega , \omega ; \bar{\omega},  \bar{\omega}) & \approx &
\Gamma_{\Lambda_0}^{(\bar{d}_{\sigma} d_{\sigma} \bar{\chi} \chi)}
( \omega , \omega ; \bar{\omega},  \bar{\omega}) = 0 ,
 \nonumber
 \\
 & &
 \label{eq:truncGamma4}
 \\
 \Gamma_{\Lambda}^{ (\bar{d}_{\uparrow} d_{\downarrow} {\chi}  )}
(\omega, \omega  -  \bar{\omega}  ,  \bar{\omega}) & \approx &
 \Gamma_{\Lambda_0}^{ (\bar{d}_{\uparrow} d_{\downarrow} {\chi}  )}
(\omega, \omega  -  \bar{\omega}  , \bar{\omega})  = 1,
 \nonumber
 \\
 & &
 \label{eq:truncGamma3a}
 \\
 \Gamma_{\Lambda}^{ ( \bar{d}_{\downarrow} d_{\uparrow} \bar{\chi}  )}
(\omega  -  \bar{\omega}  ,  \omega, \bar{\omega} ) & \approx &
 \Gamma_{\Lambda_0}^{ ( \bar{d}_{\downarrow} d_{\uparrow} \bar{\chi}  )}
(\omega  -  \bar{\omega}  ,  \omega , \bar{\omega} )
  = 1,
 \nonumber
 \\
 & &
 \label{eq:truncGamma3b}
 \end{eqnarray}
\end{subequations}

\begin{figure}[tb]
  \centering
\epsfig{file=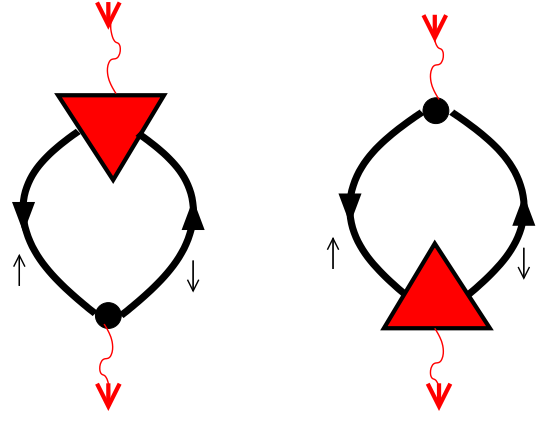,width=70mm}
\vspace{5mm}
  \vspace{-4mm}
  \caption{%
(Color online)
Graphical representation of the skeleton equation (\ref{eq:skeleton})
relating the flowing spin-flip susceptibility
$\Pi_{\Lambda}^{\bot } ( i \bar{\omega} )$ to the exact 
fermionic Green functions and the three-legged 
boson-fermion vertices. The two diagrams are equivalent.
}
\label{fig:skeleton}
\end{figure}

To close our system of flow equations, we still need an additional equation
for the flowing spin-flip susceptibility $\Pi^{\bot}_{\Lambda} ( i \bar{\omega} )$,
which in our cutoff scheme involves the pure boson vertex with four external 
legs \cite{Ledowski07}. Fortunately, we can avoid the explicit analysis of this equation
by using the Dyson-Schwinger equation for our mixed boson-fermion theory, which
implies an exact skeleton equation, relating the flowing 
 spin-flip susceptibility to the flowing fermionic Green function 
$G^{\sigma}_{\Lambda} ( i \omega )$ and the
flowing three-legged boson-fermion vertices.
Using the same method as in Appendix~B of Ref.~[\onlinecite{Schuetz05}], we 
obtain the skeleton equation
\begin{eqnarray}
 \Pi_{\Lambda}^{\bot } ( i \bar{\omega} ) & = &  - \int_{\omega}
 G_{\Lambda}^{\uparrow} ( i \omega ) G^{{\downarrow}}_{\Lambda} 
( i \omega - i \bar{\omega} )
 \nonumber
 \\
& &  \times
 \Gamma^{(\bar{d}_{\uparrow} d_{\downarrow} {\chi} )}_{\Lambda} 
(  \omega , \omega -  \bar{\omega} ,   \bar{\omega} )
\nonumber
 \\
 & = &  - \int_{\omega}
 G_{\Lambda}^{\uparrow} ( i \omega ) G^{{\downarrow}}_{\Lambda} 
( i \omega - i \bar{\omega} )
 \nonumber
 \\
& &  \times
 \Gamma^{(\bar{d}_{\downarrow} d_{\uparrow} \bar{\chi} )}_{\Lambda} 
(  \omega -  \bar{\omega} ,  \omega ,  \bar{\omega} ) .
 \label{eq:skeleton}
 \end{eqnarray}
A graphical representation of this equation is shown in Fig.~\ref{fig:skeleton}.
Note that Eq.~(\ref{eq:vertexsym}) guarantees that the two lines in
Eq.~(\ref{eq:skeleton}) are indeed identical.

For simplicity, let us focus now on the non-magnetic case, 
where all correlation functions are spin-independent and we may
omit the spin labels.
Given our approximation (\ref{eq:truncGamma3a}), (\ref{eq:truncGamma3b}),
the skeleton equation (\ref{eq:skeleton}) then reduces to
\begin{eqnarray}
 \Pi_{\Lambda}^{\bot } ( i \bar{\omega} )  & = &  - \int_{\omega}
 G_{\Lambda} ( i \omega ) G_{\Lambda} 
( i \omega - i \bar{\omega} ) ,
 \label{eq:skeleton2}
\end{eqnarray}
while the FRG flow equation (\ref{eq:selfflow}) simplifies to
 \begin{equation}
\partial_{\Lambda} \Sigma_{\Lambda} (i\omega)  = 
 \int_{\bar{\omega}} \dot{F}^{\bot}_{\Lambda} (i\bar{\omega})
G_{\Lambda} (i\omega-i  \bar{\omega})  .
 \label{eq:selfflow22}
 \end{equation}
Eqs.~(\ref{eq:skeleton2}) and (\ref{eq:selfflow22}) form a closed system of
integro-differential equations for the flowing self-energy 
$\Sigma_{\Lambda} ( i \omega )$ 
of the $d$-electrons. Recall that  Eq.~(\ref{eq:selfflow22}) 
depends implicitly on the flowing spin-flip susceptibility
via Eqs.~(\ref{eq:Fbot}) and (\ref{eq:dotFbot}).
Anticipating that there is no spontaneous magnetism,
the initial condition (\ref{eq:Sigmainitial}) for the fermionic self-energy at scale $\Lambda = \Lambda_0$
is simply 
 \begin{equation}
 \Sigma_{\Lambda_0} ( i \omega ) =  \frac{Un}{2} .
 \end{equation}
In the particle-hole symmetric case where $E_d - \mu = -U/2$ and $n=1$ this cancels precisely the energy $\xi_0^\sigma = E_d - \mu$
in the  Hartree-Fock $d$-electron propagator
$G_{\Lambda_0} ( i \omega )$, which  in the wide-band limit
is therefore given by
 \begin{equation}
 G_{\Lambda_0} ( i \omega ) = \frac{1}{ i \omega  + i \Delta {\rm sgn} \, \omega }.
 \label{eq:Ginitph}
 \end{equation}
We next show that
the solution of Eqs.~(\ref{eq:skeleton2}), (\ref{eq:selfflow22})
 does not suffer from the
Stoner instability and correctly predicts  Fermi liquid 
behavior \cite{Hewson93} for arbitrary  $U$.
We shall restrict ourselves to the particle-hole symmetric case from now on.

\subsection{Low-energy truncation}
\label{subsec:lowenergy}

The coupled system of integro-differential equations
given by Eqs.~(\ref{eq:Fbot}), (\ref{eq:dotFbot}), (\ref{eq:skeleton2}), (\ref{eq:selfflow22})
can be solved numerically for arbitrary frequencies, but this is beyond the
scope of this work.
Here, we shall focus on the low-frequency range $ | \omega | \lesssim \Delta$.
In this regime, it is reasonable to replace
the flowing $d$-electron propagator appearing
on the right-hand sides of Eqs.~(\ref{eq:skeleton2}), (\ref{eq:selfflow22})
by the low-frequency Fermi liquid form
 \begin{eqnarray}
 G_{\Lambda} ( i \omega ) & = &
 \frac{Z_l}{ i \omega + i \Delta_l {\rm sgn}\, \omega    },
 \label{eq:GFL}
\end{eqnarray}
where the flowing wave function renormalization factor is defined by
\begin{equation}
 Z_l = \frac{1}{ 1 - \left. \frac{\partial  \Sigma_{\Lambda} 
 ( i \omega ) }{ \partial ( i \omega)} \right|_{ \omega =0} },
 \end{equation}
and the renormalized flowing hybridization is
 \begin{equation}
 \Delta_l = Z_l \Delta.
 \end{equation}
We consider $Z_l$ and $\Delta_l$ to be functions
of the logarithmic flow
parameter $l = - \ln ( \Lambda / \Lambda_0 )$.
The approximation ({\ref{eq:GFL}) is certainly not sufficient at high frequencies, so that
we cannot recover in this way the high-energy Hubbard peaks at strong coupling
which are known to appear at an energy scale of order $U/2$.

Given the approximation (\ref{eq:GFL}), the frequency integration
in Eq.~(\ref{eq:skeleton2}) is easily carried out analytically, resulting in
\begin{eqnarray}
 \Pi_{\Lambda}^{\bot} ( i \bar{\omega} )  & = & 
\frac{ Z_l^2}{\pi  | \bar{\omega} | }
 \frac{ \ln \left( 1 + \frac{ | \bar{\omega} | }{\Delta_l} \right) }{
 \left[ 1 +  \frac{ | \bar{\omega} | }{2 \Delta_l}  \right] }
= \frac{ Z_l}{\pi \Delta} f \left( \frac{ | \bar{\omega} | }{\Delta_l} \right),
 \hspace{7mm}
 \label{eq:Pibotres}
 \end{eqnarray}
with
 \begin{equation}
 f ( x ) = \frac{ \ln ( 1 + x )}{ x ( 1 + x/2)}.
 \label{eq:fxdef}
 \end{equation}
Introducing the dimensionless coupling constants
 \begin{eqnarray}
 u_l & = & Z_l  u_0  =   \frac{ Z_l U}{ \pi \Delta } =    \frac{ Z^2_l U}{ \pi \Delta_l } ,
\label{eq:uldef} 
\\
 \lambda_l & = & \frac{\Lambda}{ Z_l \Delta } =   \frac{\Lambda}{  \Delta_l },
 \label{eq:lambdaldef} 
\end{eqnarray}
the single-scale propagator defined in (\ref{eq:dotFbot}) can then be written as
 \begin{eqnarray}
\dot{F}^{\bot}_{\Lambda} (i\bar{\omega}) & = &
 - \frac{ \pi}{ Z_l^2 } 
 \frac{ \Theta ( \Lambda - | \bar{\omega} | ) }{
 \left[ \frac{1}{{u}_l} + \lambda_l - \frac{ | \bar{\omega} |}{\Delta_l} -
  f (  \frac{ | \bar{\omega} |}{\Delta_l} ) \right]^2 }.
 \hspace{5mm}
 \label{eq:dotFtran}
 \end{eqnarray}
Actually, to be consistent 
with our approximation (\ref{eq:GFL}) which is based on the expansion of
the fermionic self-energy to linear order in  frequency, we should also expand the
bosonic self-energy $\Pi_\Lambda^{\bot} ( i \bar{\omega} )$ to linear order
in $\bar{\omega}$. Using $f (x ) = 1 - x  + \mathcal{O} ( x ^2)$, we have
\begin{eqnarray}
 \pi \Delta \Pi_\Lambda^{\bot} ( i \bar{\omega} )
 & \approx &  Z_l - \frac{ |\bar{\omega} |}{\Delta} + \mathcal{O} ( \bar{\omega}^2 ).
 \label{eq:Piflowapprox} 
\end{eqnarray}
As compared with the non-interacting susceptibility
(\ref{eq:Piflipsmall}), the leading constant term in Eq.~(\ref{eq:Piflowapprox})
is reduced by the wave function renormalization factor, while the
linear term is not renormalized.
In this approximation, our single-scale propagator
 (\ref{eq:dotFtran}) simplifies to
\begin{eqnarray}
\dot{F}^{\bot}_{\Lambda} (i\bar{\omega}) & = &
 - \frac{ \pi}{ Z_l^2 } 
 \frac{ \Theta ( \Lambda - | \bar{\omega} | ) }{
 \left[ \frac{1}{{u}_l} + \lambda_l - 1 \right]^2 },
 \label{eq:dotFtran2}
 \end{eqnarray}
which by construction of the regulator function given in Eqs.~(\ref{eq:RRdef}) and (\ref{eq:Litim}) is constant for frequencies below the running cutoff $\Lambda$.
To determine $Z_l$, we calculate the flowing anomalous dimension
$\eta_l$, which is  directly related to the 
derivative of the self-energy with respect to the flow parameter,
 \begin{equation}
 \eta_l = - \partial_l \ln Z_l = Z_l \Lambda \lim_{ \omega \rightarrow 0}
 \frac{\partial}{\partial ( i \omega )} \partial_{\Lambda}
 \Sigma_{\Lambda} ( i \omega ). 
\end{equation}
Substituting Eqs.~(\ref{eq:selfflow22}) and (\ref{eq:GFL}),
we obtain
 \begin{equation}
\eta_l = Z_l^2 \Lambda \lim_{ \omega \rightarrow 0}
 \frac{\partial}{\partial ( i \omega )}
  \int_{\bar{\omega}}
\frac{\dot{F}^{\bot}_{\Lambda} (i\bar{\omega})    }{i \omega - i \bar{\omega} 
 + i \Delta_l {\rm sgn}  (\omega - \bar{\omega} )  } .
 \label{eq:etalflowsing}
 \end{equation} 
If we na\"ively interchange the order of integration and differentiation,
we encounter an ambiguous expression of the form
 \begin{eqnarray}
& & \lim_{ \omega \rightarrow 0}
  \frac{\partial}{\partial ( i \omega )} 
 \left[ \frac{1}{i \omega - i \bar{\omega}  
 + i \Delta_l {\rm sgn}  (\omega - \bar{\omega} )  }
 \right] 
 \nonumber
 \\
& = & 
  \frac{1}{ [  \bar{\omega}  + \Delta_l  {\rm sgn}\, \bar{\omega}  ]^2 }
+ \frac{ 2 \Delta_l\delta (  \bar{\omega} )}{ [   
 \Delta_l ( 2 \Theta (  \bar{\omega}) -1 )  ]^2},
 \label{eq:singlim}
 \end{eqnarray}
where we have written $ {\rm sgn}\, \bar{\omega} = 2 \Theta (\bar{\omega} ) -1$.
As pointed out by Morris~\cite{Morris94},  one should interpret the product
of the delta function $\delta (x)$ with  any function  $f ( \Theta (x))$
of  the step function as
 \begin{equation}
 \delta ( x ) f ( \Theta ( x ) ) = \delta (x) \int_0^1 dt f (t).
\end{equation}
Using this relation, Eq.~(\ref{eq:singlim}) reduces to
 \begin{eqnarray}
& & \lim_{ \omega \rightarrow 0}
  \frac{\partial}{\partial ( i \omega )} 
 \left[ \frac{1}{i \omega - i \bar{\omega}  
 + i \Delta_l {\rm sgn}  (\omega - \bar{\omega} )  }
 \right] 
 \nonumber
 \\
& = & 
  \frac{1}{ [  | \bar{\omega} |  + \Delta_l  ]^2 }
-  \frac{ 2 \delta (  \bar{\omega} )}{   
 \Delta_l},
 \label{eq:singlim3}
 \end{eqnarray}
so that Eq.~(\ref{eq:etalflowsing}) takes the form
 \begin{equation}
 \eta_l = - \frac{Z_l \Lambda}{\pi \Delta} \dot{F}^{\bot}_{\Lambda} (i0)
  + Z_l^2 \Lambda \int_{\bar{\omega}} \frac{ \dot{F}^{\bot}_{\Lambda} 
 (i \bar{\omega})}{ ( | \bar{\omega} | + \Delta_l )^2 }.
 \label{eq:etalflow}
 \end{equation}
With the single-scale propagator given in Eq.~(\ref{eq:dotFtran2}), the
frequency integration is trivial and we finally obtain
from Eq.~(\ref{eq:etalflowsing})
for the flowing anomalous dimension
 \begin{eqnarray}
 \eta_l & = &   \frac{ {\lambda}_l}{ [1 +  {\lambda}_l] [  \frac{1}{u_l} + \lambda_l -1 ]^2 }.
 \label{eq:etares2} 
\end{eqnarray}
The right-hand side of this expression depends 
on $Z_l$ via the flowing couplings $u_l$ and $\lambda_l$ defined
in Eqs.~(\ref{eq:uldef}) and (\ref{eq:lambdaldef}) so that
 \begin{equation}
 \partial_l Z_l = - \eta_l Z_l
 \label{eq:Zflow}
 \end{equation}
is an ordinary differential equation for the flowing wave function
renormalization factor $Z_l$, which is easily solved numerically.
The result for the flowing $Z_l$ is shown in Fig.~\ref{fig:Zflow}.
\begin{figure}[tb]    
   \centering
 \epsfig{file=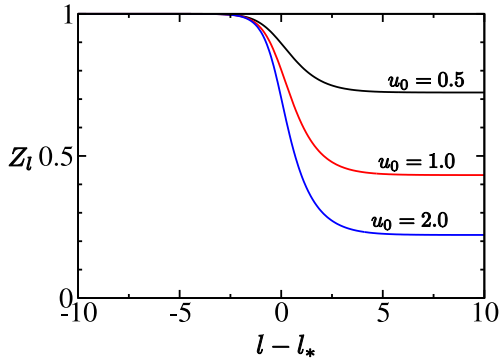,width=70mm}
  \caption{%
Numerical solution of Eq.~(\ref{eq:Zflow})
for different values of the bare coupling $u_0 = U /( \pi \Delta)$.
Here, $l_{\ast}$
is the scale where the running cutoff $\Lambda$ is equal to the hybridization
$\Delta$, i.e.,  $ \Lambda_0 e^{-l_{\ast}} = \Delta$.
}
    \label{fig:Zflow}
  \end{figure}
Obviously, for $l \rightarrow \infty$
the wave function renormalization factor approaches a finite limit,
$
Z = \lim_{l \rightarrow \infty} Z_l $,
which we show in Fig.~\ref{fig:Zofg} as a function of the bare coupling $u_0$.
\begin{figure}[tb]    
   \centering
 \epsfig{file=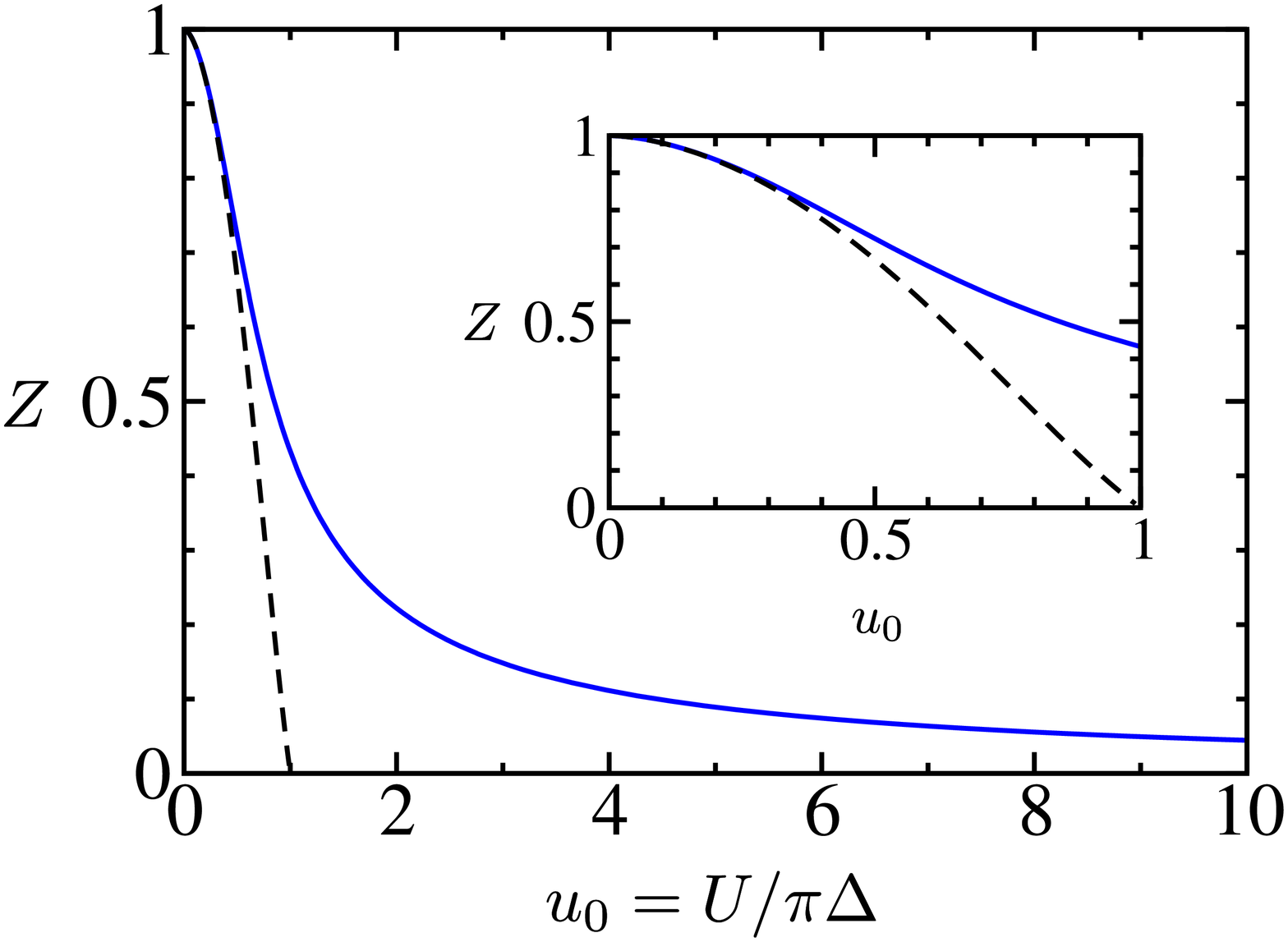,width=70mm}
  \caption{%
The solid line is the wave function renormalization factor
$Z = \lim_{ l \rightarrow \infty} Z_l$ as a function
of the bare coupling $u_0$ obtained from the
numerical solution of our truncated flow equation (\ref{eq:Zflow}).
For comparison, we show as the dashed line the result of the ladder approximation.
}
    \label{fig:Zofg}
  \end{figure}
In contrast to the ladder approximation discussed in
Sec.~\ref{sec:Ladder} we now obtain a finite $Z$ for all values
of the interaction, so that the fluctuations included
in our simple FRG approach are sufficient to
remove  the unphysical Stoner instability.
On the other hand, quantitatively our truncation of the exact FRG 
flow equations does not reproduce the correct strong-coupling behavior
of the quasiparticle residue, which is known to exhibit 
the same
exponential suppression as the Kondo temperature for $u_0 \to \infty$ (see Ref.~[\onlinecite{Tsvelick83}]
),
 \begin{equation}
Z_{\rm exact} \sim \sqrt{\frac{8 u_0}{\pi}} \exp [ - \pi^2 u_0/8].
 \label{eq:Zexact}
\end{equation}
From the numerical solution of our flow equation given in 
Eqs.~(\ref{eq:etares2}), (\ref{eq:Zflow}) we find asymptotically
\begin{equation}
 Z \sim \frac{0.445}{u_0}, \; \; \; u_0 \gg 1.
 \label{eq:Zstrong}
 \end{equation}
The fact that this is much larger than the exact result 
(\ref{eq:Zexact}) indicates that
our approach underestimates the strength of fluctuations.
In the following section we shall therefore include the longitudinal spin fluctuation channel
and show that it indeed further suppresses  the value of $Z$, 
although within our approximations we are unable to recover 
the exponential suppression described by Eq.~(\ref{eq:Zexact}).

\section{FRG with partial bosonization of transverse and longitudinal spin fluctuations}
\label{sec:twospin}

\subsection{Multi-channel Hubbard-Stratonovich transformation}

Using the antisymmetry of the Grassmann-fields, the
local Hubbard interaction can be written in infinitely many equivalent ways,
such as
\begin{equation}
U n_{\uparrow} ( \tau )    n_{\downarrow}  ( \tau )  =
 - \frac{U^{\parallel}}{2}  m^2 ( \tau ) - U^{\bot}
 \bar{s} ( \tau ) s( \tau ),
 \label{eq:decomp}
 \end{equation} 
where
 \begin{equation}
  U^{\parallel} + U^{\bot} = U .
 \label{eq:UU}
 \end{equation}
Here,
$ n_\sigma ( \tau ) = \bar{d}_{\sigma} ( \tau ) d_{\sigma} ( \tau )$ and the composite fields
$m ( \tau ) = \sum_{\sigma} \sigma  n_{\sigma} ( \tau )$, 
$\bar{s} ( \tau ) = \bar{d}_{\uparrow} ( \tau ) {d}_{\downarrow} ( \tau )$, 
and $s ( \tau ) = \bar{d}_{\downarrow} ( \tau ) d_{\uparrow} ( \tau )$ represent the longitudinal and 
transverse spin components, see also Eq.~(\ref{eq:sflipdef}).
Note that by construction of the decomposition (\ref{eq:decomp}) we have included both longitudinal and transverse spin fluctuations, but no charge fluctuations.
In Sec.~\ref{sec:FRG1} we have satisfied
Eq.~(\ref{eq:decomp}) by setting
$U^{\parallel} =0$ and $U^{\bot} =U$, which is the natural choice  
if one is interested in bosonizing
the transverse spin fluctuations.
Alternatively, one could set $U^{\parallel} = U$ and $U^{\bot} =0$, so that
 longitudinal spin fluctuations can be
introduced via a suitable Hubbard-Stratonovich field.
The ambiguity  associated
with Eqs.~(\ref{eq:decomp}) and (\ref{eq:UU})
has been discussed for many decades in the 
literature \cite{Hamann69,Wang69,Castellani79,Schulz90,Macedo91,Dupuis02}.
Depending on the physical problem of interest, certain special choices
of $U^{\parallel}$ and $U^{\bot}$ can be advantageous \cite{Macedo91}.
At this point, we simply leave the precise values
of $U^{\parallel}$ and $U^{\bot}$ unspecified but assume that
both are positive and satisfy Eq.~(\ref{eq:UU}).
Later we shall show that in the strong coupling regime of the particle-hole
symmetric AIM the optimal choice is $ U^{ \parallel} = U/3$ and
$U^{\bot} = 2 U /3$. 
In this case the decomposition (\ref{eq:decomp})
is manifestly spin-rotational invariant and can be written as
 \begin{equation}
U n_{\uparrow} ( \tau )    n_{\downarrow}  ( \tau )  =
 - \frac{U}{6}  \left( \vec{s} ( \tau ) \right)^2 ,
 \label{eq:HSrotinv}
 \end{equation}
where the composite vector field 
$\vec{s} ( \tau ) = d^{\dagger} ( \tau )  \vec{\sigma} d ( \tau )$
represents the spin vector.
Here, $ d^{\dagger} ( \tau ) =  [ \bar{d}_{\uparrow} ( \tau ) ,  
\bar{d}_{\downarrow} ( \tau ) ]$  and $\vec{\sigma}$ is the usual
matrix vector of Pauli matrices.

Starting from the representation (\ref{eq:decomp}), we decouple
the interaction in the 
longitudinal spin channel using a real Hubbard-Stratonovich field
$\eta$ and in the transverse spin channel using 
the complex Hubbard-Stratonovich fields $\chi$ and $\bar{\chi}$ introduced
in Sec.~\ref{subsec:HStrans}.
The partition function can then be written as in Eq.~(\ref{eq:Zratio2}), where 
$\Phi = [ d_{  \uparrow} , \bar{d}_{ \uparrow} ,
 d_{  \downarrow} , \bar{d}_{ \downarrow} ,  
\eta , \chi , \bar{\chi}]$ is now a seven-component super-field.
For the Gaussian part of the action we then obtain instead of 
Eq.~(\ref{eq:S0Phi})
 \begin{align}
 S_0 [ \Phi ] =
&- \int_{ \omega} \sum_{  \sigma  }
 \bigl[ {G}_0^{\sigma} ( i \omega ) 
 \bigr]^{-1} \bar{d}_{ \omega \sigma}  {d}_{ \omega \sigma}    
 \nonumber
 \\
{} &+  
 \int_{ \bar{\omega}} 
(U^{\bot})^{-1} \bar{\chi}_{  \bar{\omega}} \chi_{  \bar{\omega}} +  
  \frac{1}{2} \int_{ \bar{\omega}} 
(U^{\parallel})^{-1} 
\eta_{ - \bar{\omega} } \eta_{ \bar{\omega} } , 
 \label{eq:S0Phimulti}
\end{align}
and the interaction part can be written as
 \begin{align}
 S_1 [ \Phi ] = &\int_{\bar{\omega}} \left[
\bar{s}_{ \bar{\omega}}  {\chi}_{ \bar{\omega}} + 
    s_{ \bar{\omega}} \bar{\chi}_{ \bar{\omega}} 
 + m_{-\bar{\omega}} \eta_{\bar{\omega}} 
 \right] 
 \nonumber
 \\
{} - &\int_{\omega } \sum_{\sigma} \delta\xi^{\sigma}  
\bar{d}_{ \omega \sigma}  {d}_{ \omega \sigma} .
\label{eq:S1PhimultiHS}
 \end{align}
The counterterm $\delta \xi^{\sigma}$
in Eq.~(\ref{eq:S1PhimultiHS})
is subtracted to correct for the inclusion of the Hartree-Fock self-energy
in the propagator
$ {G}_0^{\sigma} ( i \omega ) $ in Eq.~(\ref{eq:S0Phimulti}).
While the transverse spin components $s_{ \bar{\omega} }$ and $\bar s_{ \bar{\omega} }$ are given by Eqs.~(\ref{eq:defs}) and (\ref{eq:defsbar}), the longitudinal spin component is given by
  \begin{equation}
m_{ \bar{\omega} } = 
\int_0^{\beta} d \tau  e^{ i \bar{\omega} \tau } m ( \tau ) = 
\int_{\omega}  \sum_{\sigma} \sigma 
 \bar{d}_{ \omega  \sigma}
 d_{ \omega + \bar{\omega} , \sigma}
 .
\end{equation}
 For simplicity, we shall focus here on the non-magnetic, 
particle-hole-symmetric case. Since in the strong coupling limit the low-energy physics is expected
to be dominated by spin fluctuations, 
we set up the FRG by
introducing a cutoff only in the bosonic fields $\eta,\chi$ and $\bar{\chi}$ associated
with spin fluctuations. Particle-hole symmetry guarantees that
the exact self-energy at zero frequency cancels the term $\epsilon_d - \mu$, so that we may 
approximate $G_0^{\sigma} ( i \omega )$ by Eq.~(\ref{eq:Ginitph}).

\subsection{FRG flow equations}

We now generalize Eq.~(\ref{eq:UR}) by introducing cutoffs 
into both types of bosonic propagators appearing
in the Gaussian action (\ref{eq:S0Phimulti}),
 \begin{eqnarray}
 ( U^{\alpha})^{-1} & \rightarrow &  (U^{\alpha})^{-1} + R^{\alpha}_{\Lambda} (  i  \bar{\omega}  )  \; ,
 \end{eqnarray}
where $ \alpha=\parallel, \bot$ labels the two types of spin fluctuations, 
and the cutoff functions are
  \begin{subequations}
 \begin{eqnarray}
  R^{\parallel} ( i \bar{\omega} ) & = & \frac{ 2 \Lambda}{\pi \Delta^2 }
 R \left( \frac{ | \bar{\omega} |}{\Lambda } \right) ,
 \\
R^{\bot} ( i \bar{\omega} ) & = & \frac{  \Lambda}{\pi \Delta^2 }
 R \left( \frac{ | \bar{\omega} |}{\Lambda } \right) ,
 \end{eqnarray}
 \end{subequations}
with $R (x )$ given in Eq.~(\ref{eq:Litim}).
From the general FRG flow equations for mixed boson-fermion theories given
in Refs.~[\onlinecite{Schuetz05,Kopietz09}]
we then obtain the following exact flow equation for the self-energy,
 \begin{widetext}
\begin{eqnarray} 
 \partial_{\Lambda} \Sigma^{\sigma}_{\Lambda}(i\omega) & = &
\int_{\bar{\omega}} 
\dot{F}^{\parallel}_{\Lambda} (i\bar{\omega})
\Gamma_{\Lambda}^{(\bar{d}_{\sigma} d_{\sigma} \eta \eta)}
( \omega , \omega ;  \bar{\omega},  - \bar{\omega}) 
+ 
\int_{\bar{\omega}} 
\dot{F}^{\bot}_{\Lambda} (i\bar{\omega})
\Gamma_{\Lambda}^{(\bar{d}_{\sigma} d_{\sigma} \bar{\chi} \chi)}
( \omega , \omega ; \bar{\omega},  \bar{\omega})
\nonumber \\
%
& + &  \int_{\bar{\omega}} \dot{F}^{\parallel}_{\Lambda} (i\bar{\omega})
G_{\Lambda}^{\sigma} (i\omega-i\bar{\omega})   
\Gamma_{\Lambda}^{ (\bar{d}_{\sigma} d_{\sigma} \eta  )}
(\omega, \omega  - \bar{\omega} ,   \bar{\omega})
\Gamma_{\Lambda}^{  (\bar{d}_{\sigma} d_{\sigma}  \eta)   }
(  \omega - \bar{\omega},  \omega , - \bar{\omega}) 
\nonumber
 \\
&  +  &
 \int_{\bar{\omega}} \dot{F}^{\bot}_{\Lambda} (i\bar{\omega})
G_{\Lambda}^{\bar{\sigma}} (i\omega - i \sigma \bar{\omega})   
\Gamma_{\Lambda}^{  (\bar{d}_{\sigma} d_{\bar{\sigma}} {\chi}_{\sigma})   }
( \omega , \omega - \sigma \bar{\omega},  \sigma \bar{\omega})
\Gamma_{\Lambda}^{ (\bar{d}_{\bar{\sigma}} d_{\sigma} {\chi}_{\bar{\sigma}}  )}
(\omega - \sigma \bar{\omega} , \omega , \sigma \bar{\omega}),
 \label{eq:selfflowlontra}
 \end{eqnarray}
\end{widetext}
which generalizes Eq.~(\ref{eq:selfflow})
and is shown graphically in Fig.~\ref{fig:selflontra}.
\begin{figure}[tb]
  \centering
\epsfig{file=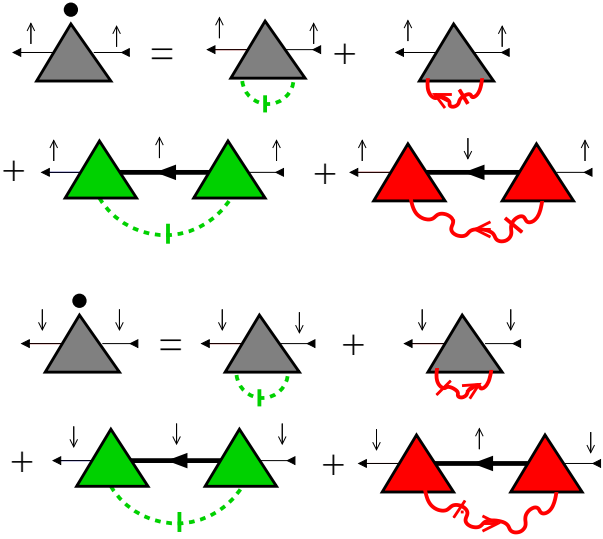,width=80mm}
\vspace{5mm}
  \vspace{-4mm}
  \caption{%
(Color online)
Graphical representation of the exact FRG flow equation (\ref{eq:selfflowlontra}) for the
self-energies $\Sigma^{\uparrow}_{\Lambda}$ and  
$\Sigma^{\downarrow}_{\Lambda}$.
The longitudinal spin-field $\eta$ is represented by dashed lines; the other symbols
are the same as in Fig.~\ref{fig:ph-selfenergy}.
}
\label{fig:selflontra}
\end{figure}
The superscripts label the different types of vertices;
for the boson-fermion vertices associated 
with the transverse spin field
we have used again the short 
notation $\chi_{\uparrow} = \chi$ and $\chi_{\downarrow} =
\bar{\chi}$, see Eq.~(\ref{eq:selfflow}).
The longitudinal and transverse single-scale propagators are
 \begin{equation}
 \dot{F}^{\alpha}_{\Lambda} ( i \bar{\omega} ) = [ - \partial_{\Lambda}
 R_{\Lambda}^{\alpha} ( i \bar{\omega}  ) ] [ F^{\alpha}_{\Lambda} ( i \bar{\omega} ) ]^2,
 \label{eq:dotFalpha} 
\end{equation}
with
\begin{eqnarray}
 F^{\alpha}_{\Lambda} ( i \bar{\omega} ) & = &  
[ (U^{\alpha})^{-1} -  \Pi^{\alpha }_{\Lambda} ( i \bar{\omega} ) +
R_{\Lambda}^{\alpha}   ( i  \bar{\omega}  ) ]^{-1},
 \label{eq:Falpha}
 \end{eqnarray}
where $\Pi^{\alpha }_{\Lambda} ( i \bar{\omega} )$ are the
 corresponding  irreducible spin susceptibilities.
As discussed in Sec.~\ref{subsec:trunc}, we use the skeleton equations (which can
be formally derived from the Dyson-Schwinger equations) to relate the flow
of the susceptibilities to the flow of the $d$-electron propagators 
and the three-legged boson-fermion vertex. The skeleton equation for the
transverse susceptibility 
$\Pi^{\bot}_{\Lambda} ( i \bar{\omega} )$ has already been discussed
in Sec.~\ref{subsec:trunc}, see Eq.~(\ref{eq:skeleton}); it can also be written as 
 \begin{eqnarray}
  \Pi_{\Lambda}^{\bot } ( i \bar{\omega} ) 
 & = &
  - \int_{\omega}
 G_{\Lambda}^{\sigma} ( i \omega ) G^{{\bar{\sigma}}}_{\Lambda} 
( i \omega - i \sigma \bar{\omega} )
\nonumber
 \\
  & & \hspace{10mm} \times 
 \Gamma^{(\bar{d}_{\sigma} d_{\bar{\sigma}} {\chi}_{\sigma} )}_{\Lambda} 
(  \omega , \omega -  \sigma \bar{\omega} ,   \bar{\omega} ).
\nonumber
 \\
 \label{eq:skeletontr}
 \end{eqnarray}
The corresponding skeleton equation for the
longitudinal spin susceptibility is
\begin{eqnarray}
 \Pi^{\parallel}_{\Lambda} ( i \bar{\omega} ) 
 & = & - \int_{\omega} \sum_{\sigma} \sigma 
G^{\sigma}_{\Lambda} ( i \omega )
 G^{\sigma}_{\Lambda} ( i \omega - i \bar{\omega})
 \nonumber
 \\
 & & \hspace{10mm} \times 
 \Gamma_{\Lambda}^{ (\bar{d}_{\sigma} d_{\sigma} \eta) } ( \omega , \omega -
 \bar{\omega}  , \bar{\omega} ).
 \label{eq:skeletonlon}
 \end{eqnarray}
To obtain a closed RG flow equation for the self-energy, we
still need flow equations for the vertices with three and four external legs
appearing in Eq.~(\ref{eq:selfflowlontra}).
As in Sec.~\ref{subsec:trunc}, we simply set all mixed boson-fermion vertices with
four external legs equal to zero, because these vertices vanish at the initial scale.
However, in contrast to the FRG with only transverse spin fluctuations
discussed in Sec.~\ref{subsec:trunc}, the renormalization of the 
three-legged boson-fermion vertices is now 
important. In the approximation where mixed boson-fermion vertices with four and more
external legs are ignored, the FRG flow equation for the 
longitudinal three-legged boson-fermion vertex is
\begin{eqnarray}
& & \partial_{\Lambda} \Gamma_{\Lambda}^{  (\bar{d}_{\sigma} d_{\sigma} \eta)   } 
( \omega + \bar{\omega} , \omega , \bar{\omega} ) 
\nonumber
 \\
 & = & 
 \int_{ \bar{\omega}^{\prime} } 
\dot{F}^{\parallel}_{\Lambda} (i\bar{\omega}^{\prime})
G_{\Lambda}^{\sigma}  (  i {\omega} + i \bar{\omega} +  i\bar{\omega}^{\prime} )
G_{\Lambda}^{\sigma}  (i \omega + i\bar{\omega}^{\prime}  )
\nonumber
\\
&    & \times
 \Gamma_{\Lambda}^{( \bar{d}_{\sigma} {d}_{\sigma} \eta )}
 ( \omega + \bar{\omega} , \omega + \bar{\omega} + \bar{\omega}^{\prime}  , 
- \bar{\omega}^{\prime} )
\nonumber
\\
&    & \times
\Gamma_{\Lambda}^{( \bar{d}_{\sigma} {d}_{\sigma} \eta )}
 ( \omega + \bar{\omega} + \bar{\omega}^{\prime}  , 
\omega + \bar{\omega}^{\prime} , \bar{\omega} )
 \nonumber
 \\
 & & \times
\Gamma_{\Lambda}^{( \bar{d}_{\sigma} {d}_{\sigma} \eta )}
 ( \omega + \bar{\omega}^{\prime} ,  {\omega}  , 
\bar{\omega}^{\prime} )
\nonumber
 \\
 & + & \int_{ \bar{\omega}^{\prime} } 
\dot{F}^{\bot}_{\Lambda} (- i \sigma \bar{\omega}^{\prime})
G_{\Lambda}^{\bar{\sigma}}  (  i {\omega} + i \bar{\omega} +  i\bar{\omega}^{\prime} )
G_{\Lambda}^{\bar{\sigma}}  (i \omega + i\bar{\omega}^{\prime}  )
\nonumber
\\
&    & \times
 \Gamma_{\Lambda}^{( \bar{d}_{\sigma} {d}_{\bar{\sigma}} \chi_{\sigma} )}
 ( \omega + \bar{\omega} , \omega + \bar{\omega} + \bar{\omega}^{\prime}  , 
- \sigma \bar{\omega}^{\prime} )
\nonumber
\\
&    & \times
\Gamma_{\Lambda}^{( \bar{d}_{\sigma} {d}_{\sigma} \eta )}
 ( \omega + \bar{\omega} + \bar{\omega}^{\prime}  , 
\omega + \bar{\omega}^{\prime} , \bar{\omega} )
 \nonumber
 \\
 & & \times
\Gamma_{\Lambda}^{( \bar{d}_{\bar{\sigma}} {d}_{\sigma} \chi_{\bar{\sigma}} )}
 ( \omega + \bar{\omega}^{\prime} ,  {\omega}  , 
- \sigma \bar{\omega}^{\prime} ) ,
 \label{eq:flowGammalon}
 \end{eqnarray}
while the corresponding flow equation for the transverse vertex reads
\begin{eqnarray}
 & & \partial_{\Lambda} \Gamma_{\Lambda}^{  (\bar{d}_{\sigma} 
 d_{\bar{\sigma}}  \chi_{\sigma})   }  ( \omega + \bar{\omega} , \omega , \sigma \bar{\omega} ) 
\nonumber
 \\
 & = & 
 \int_{ \bar{\omega}^{\prime} } 
\dot{F}^{\parallel}_{\Lambda} (i\bar{\omega}^{\prime})
G_{\Lambda}^{\sigma}  (  i {\omega} + i \bar{\omega} +  i\bar{\omega}^{\prime} )
G_{\Lambda}^{\bar{\sigma}}  (i \omega + i\bar{\omega}^{\prime}  )
\nonumber
\\
&    & \times
 \Gamma_{\Lambda}^{( \bar{d}_{\sigma} {d}_{\sigma} \eta )}
 ( \omega + \bar{\omega} , \omega + \bar{\omega} + \bar{\omega}^{\prime}  , 
- \bar{\omega}^{\prime} )
\nonumber
\\
&    & \times
\Gamma_{\Lambda}^{( \bar{d}_{\sigma} {d}_{\bar{\sigma}} \chi_{\sigma} )}
 ( \omega + \bar{\omega} + \bar{\omega}^{\prime}  , 
\omega + \bar{\omega}^{\prime} , \sigma \bar{\omega} )
 \nonumber
 \\
 & & \times
\Gamma_{\Lambda}^{( \bar{d}_{\bar{\sigma}} {d}_{\bar{\sigma}} \eta )}
 ( \omega + \bar{\omega}^{\prime} ,  {\omega}  , 
\bar{\omega}^{\prime} ).
 \label{eq:flowGammatrans}
 \end{eqnarray}
Graphical representations of Eqs.~(\ref{eq:flowGammalon}) and
(\ref{eq:flowGammatrans}) are shown in Figs.~\ref{fig:flowGammalon} and
\ref{fig:flowGammatrans}.
From  $S_1 [ \Phi ]$ 
in Eq.~(\ref{eq:S1PhimultiHS}) we see that
the initial conditions for the three-legged vertices at scale $\Lambda = \Lambda_0$ are
 \begin{subequations}
 \begin{eqnarray}
\Gamma_{\Lambda_0}^{ (\bar{d}_{\sigma} d_{\sigma} \eta) } 
( \omega + \bar{\omega},  \omega   , \bar{\omega} ) & = & \sigma ,
 \label{eq:Gammaloninit}
\\
 \Gamma^{(\bar{d}_{\sigma} d_{\bar{\sigma}} {\chi}_{\sigma} )}_{\Lambda_0} 
(  \omega + \bar{\omega} , \omega ,  \sigma \bar{\omega}  ) & = & 1.
 \label{eq:Gammatrainit}
\end{eqnarray}
\end{subequations}
\begin{figure}[tb]
  \centering
\epsfig{file=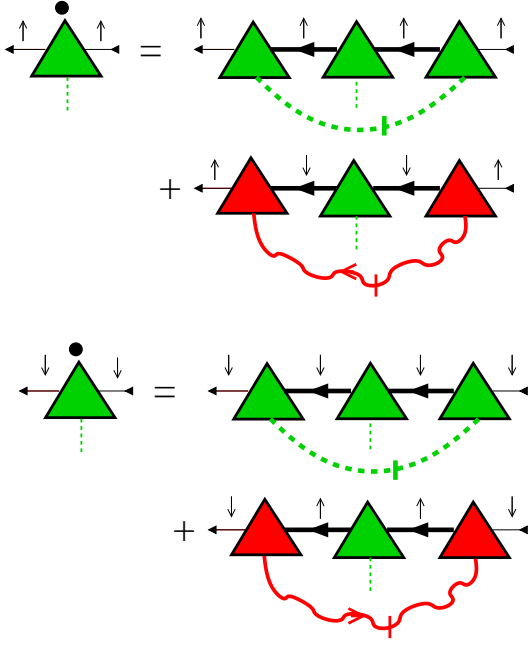,width=70mm}
\vspace{5mm}
  \vspace{-4mm}
  \caption{%
(Color online)
Graphical representation of the FRG flow equation (\ref{eq:flowGammalon})
for the longitudinal boson-fermion vertices
 $\Gamma_{\Lambda}^{  ( \bar{d}_{\uparrow} 
 d_{\uparrow}  \eta )   }$ (upper graph) and
$\Gamma_{\Lambda}^{  ( \bar{d}_{\downarrow} 
 d_{\downarrow}  \eta )   }$ (lower graph).
}
\label{fig:flowGammalon}
\end{figure}
\begin{figure}[tb]
  \centering
\epsfig{file=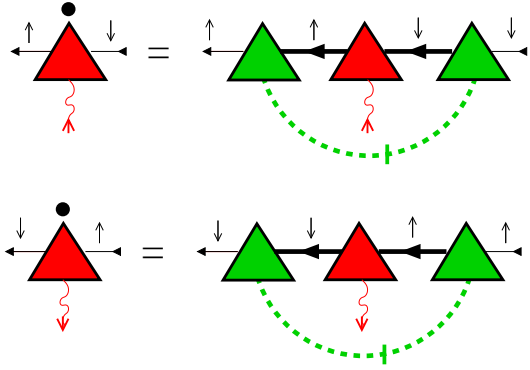,width=70mm}
\vspace{5mm}
  \vspace{-4mm}
  \caption{%
(Color online)
Graphical representation of the FRG flow equation (\ref{eq:flowGammatrans})
for the transverse boson-fermion vertices
 $\Gamma_{\Lambda}^{  ( \bar{d}_{\uparrow} 
 d_{\downarrow}  \chi )   }$ (upper graph) and
 $\Gamma_{\Lambda}^{  (
\bar{d}_{\downarrow}  d_{\uparrow}  \bar{\chi} )   }$ (lower graph).
}
\label{fig:flowGammatrans}
\end{figure}

\subsection{Low-energy truncation}
To make further progress, we now neglect the frequency-dependence of the three-legged boson-fermion 
vertices, setting
 \begin{eqnarray}
\Gamma_{\Lambda}^{ (\bar{d}_{\sigma} d_{\sigma} \eta) } ( \omega + \bar{\omega},  
\omega   , \bar{\omega} ) & \approx & \sigma \gamma^{\parallel}_{\Lambda},
 \\
\Gamma_{\Lambda}^{ (\bar{d}_{\sigma} d_{\bar{\sigma}}  \chi_{\sigma} ) } 
( \omega + \bar{\omega},  \omega   , \sigma \bar{\omega} ) & \approx &  
\gamma^{\bot}_{\Lambda}
 .
 \end{eqnarray}
Keeping in mind that we have also neglected all vertices involving more than three external
legs, our exact FRG flow equation (\ref{eq:selfflowlontra}) for the self-energy
 reduces to
\begin{eqnarray} 
 \partial_{\Lambda} \Sigma^{\sigma}_{\Lambda}(i\omega) & = &
 ( \gamma_\Lambda^{\parallel})^2 \int_{ \bar{\omega}}   
\dot{F}^{\parallel}_{\Lambda} (i\bar{\omega})
 G_{\Lambda}^{\sigma}  (  i \omega - i \bar{\omega}  ) 
 \nonumber
 \\
 & + & ( \gamma_\Lambda^{\bot})^2 \int_{ \bar{\omega}}   
\dot{F}^{\bot}_{\Lambda} (i\bar{\omega})
 G_{\Lambda}^{\bar{\sigma}}  (  i \omega - i \sigma \bar{\omega}  ) .
 \hspace{7mm}
\label{eq:selfflow2} 
\end{eqnarray}
Moreover,  the approximate FRG flow equation (\ref{eq:flowGammalon}) 
for the longitudinal spin-fermion vertex reduces to
\begin{eqnarray}
 \partial_{\Lambda} \gamma_{\Lambda}^{\parallel} 
 & = & ( \gamma_{\Lambda}^{\parallel})^3 
 \int_{ \bar{\omega} } 
\dot{F}^{\parallel}_{\Lambda} (i\bar{\omega})
[ G_{\Lambda}^{\uparrow}  (  i \bar{\omega}  ) ]^2
\nonumber
\\
 & -  &   \gamma_{\Lambda}^{\parallel}   ( \gamma_{\Lambda}^{\bot})^2 
 \int_{ \bar{\omega} } 
\dot{F}^{\bot}_{\Lambda} (i\bar{\omega})
[ G_{\Lambda}^{\downarrow}  (  - i \bar{\omega}  ) ]^2
\; ,
 \label{eq:flowGammalon2}
 \end{eqnarray}
while the FRG equation  (\ref{eq:flowGammatrans}) for the transverse
spin-fermion vertex becomes
\begin{eqnarray}
 \partial_{\Lambda} \gamma_{\Lambda}^{\bot} 
 & = & -    \gamma_{\Lambda}^{\bot}   ( \gamma_{\Lambda}^{\parallel})^2 
 \int_{ \bar{\omega} } 
\dot{F}^{\parallel}_{\Lambda} (i\bar{\omega})
G_{\Lambda}^{\uparrow}  (   i \bar{\omega}  ) 
 G_{\Lambda}^{\downarrow}  (   i \bar{\omega}  ) 
\; . \hspace{10mm}
 \label{eq:flowGammatra2}
 \end{eqnarray}
Within the same approximation, the skeleton equations (\ref{eq:skeletontr}) and
(\ref{eq:skeletonlon}) for the spin susceptibilities are
\begin{eqnarray}
 \Pi^{\parallel}_{\Lambda} ( i \bar{\omega} ) 
 & = & - \gamma_{\Lambda}^{\parallel} \int_{\omega} \sum_{\sigma} 
G^{\sigma}_{\Lambda} ( i \omega )
 G^{\sigma}_{\Lambda} ( i \omega - i \bar{\omega}),
 \hspace{5mm}
 \label{eq:skeletonlonapprox}
 \\
  \Pi_{\Lambda}^{\bot } ( i \bar{\omega} ) 
 & = &
  - \gamma_{\Lambda}^{\bot} \int_{\omega}
 G_{\Lambda}^{\sigma} ( i \omega ) G^{{\bar{\sigma}}}_{\Lambda} 
( i \omega - i \sigma \bar{\omega} ).
 \label{eq:skeletontrapprox}
 \end{eqnarray}
Eqs.~(\ref{eq:selfflow2})--(\ref{eq:skeletontrapprox}) form a closed system of
integro-differential equations for the frequency-dependent self-energy
$\Sigma_{\Lambda}^{\sigma} ( i \omega )$ and
the frequency-independent parts $\gamma_\Lambda^{\parallel}$ and 
$\gamma_\Lambda^{\bot}$ of the three-legged boson-fermion vertices.
In order to make progress analytically, let us now approximate 
the flowing $d$-electron Green function on the right-hand sides
of the flow equation by its low-energy Fermi liquid form 
(\ref{eq:GFL}). Then the integrations in Eqs.~(\ref{eq:skeletonlonapprox}), (\ref{eq:skeletontrapprox}) can be performed analytically.
For simplicity, we focus again on the non-magnetic particle-hole symmetric case
and omit the spin-labels. We then obtain
 \begin{eqnarray}
  \Pi_{\Lambda}^{\parallel} ( i \bar{\omega} ) & = &  \frac{ 2 Z_l }{\pi \Delta } 
   \gamma_l^{\parallel} f   
\left( \frac{ |\bar{\omega} | }{\Delta_l}   \right),
 \label{eq:Pilonapprox}
 \\
 \Pi^{\bot}_{\Lambda} ( i \bar{\omega} ) & = & \frac{Z_l}{\pi \Delta } \gamma_l^{\bot}
 f \left( \frac{ |\bar{\omega} | }{\Delta_l} \right),
 \label{eq:Pibotapprox}
 \end{eqnarray}
where the function $f( x )$ is given in Eq.~(\ref{eq:fxdef}).
The bosonic  single-scale propagators can then be written as 
\begin{eqnarray}
 \dot{F}^{\parallel}_{\Lambda} (i\bar{\omega}) & = &
 - \frac{ \pi}{2 Z_l^2} 
 \frac{ \Theta ( \Lambda - | \bar{\omega} | ) }{
 \left[ \frac{1}{ {u}_l^{\parallel}} + \lambda_l - \frac{ | \bar{\omega} |}{\Delta_l} -
 \gamma_l^{\parallel} f (  \frac{ | \bar{\omega} |}{\Delta_l}  ) \right]^2 },
 \nonumber
 \label{eq:dotFlon3}
 \\
 & &
 \\
\dot{F}^{\bot}_{\Lambda} (i\bar{\omega}) & = &
 - \frac{ \pi}{ Z_l^2 } 
 \frac{ \Theta ( \Lambda - | \bar{\omega} | ) }{
 \left[ \frac{1}{{u}_l^{\bot}} + \lambda_l - \frac{ | \bar{\omega} |}{\Delta_l} -
 \gamma_l^{\bot} f (  \frac{ | \bar{\omega} |}{\Delta_l}  ) \right]^2 }.
 \nonumber
 \\
 & &
 \label{eq:dotFtran3}
 \end{eqnarray}
Here, we have introduced again the notation $\lambda_l = \Lambda / \Delta_l = \Lambda /(Z_l \Delta )$, and
the running interaction constants ${u}_l^{\parallel}$ and ${u}_l^{\bot}$
are defined by
 \begin{equation}
 {u}_l^{\parallel} = Z_l  \frac{ 2 U^{\parallel}}{ \pi \Delta } \; \; , \; \;
 {u}_l^{\bot}  =  Z_l \frac{  U^{\bot}}{ \pi \Delta }.
 \label{eq:ulbot}
\end{equation}
In terms of the dimensionless bare couplings 
$u_0^{\parallel}$ and $u_0^{\bot}$
the condition (\ref{eq:UU})  reads 
\begin{equation}
u_0^{\parallel}/2 + u_0^{\bot} = u_0 .
\end{equation}
As discussed in Sec.~\ref{subsec:lowenergy},
for consistency
we expand the function $f ( | \bar{\omega} | / \Delta_l )$
in Eqs.~(\ref{eq:Pilonapprox}) and (\ref{eq:Pibotapprox}) to linear order
in frequency as in Eq.~(\ref{eq:Piflowapprox}). 
Then the frequency integrations in Eqs.~(\ref{eq:selfflow2}), (\ref{eq:flowGammalon2}),
(\ref{eq:flowGammatra2}) can be performed analytically and we obtain the following
system of RG flow equations for the three running couplings
$Z_l$, $\gamma_l^{\parallel}$, and $\gamma_l^{\bot}$,
 \begin{align}
 \frac{\partial_l Z_l}{Z_l}   &=  -  \eta_l^{\parallel} - \eta_l^{\bot}   ,
 \label{eq:Zlogflow}
 \\
 \frac{\partial_l \gamma_l^{\parallel}}{\gamma_l^{\parallel}  }
  &=  - \frac12 A_l^{\parallel} + A_l^{\bot}  ,
 \label{eq:gammatralogflow}
 \\
 \frac{\partial_l \gamma_l^{\bot}}{\gamma_l^{\bot} } &=  \frac12 A_l^{\parallel},
\label{eq:gammalonlogflow}
 \end{align}
where
\begin{align}
 \eta_l^{\parallel}  &= \frac{ (\gamma_l^{\parallel} )^2 \lambda_l  }{2 
 \bigl[ 1/ {u}_l^{\parallel}  + \lambda_l - 
 \gamma_l^{\parallel}  \bigr]^{2}}   - \frac12 A_l^{\parallel},
 \\
 \eta_l^{\bot}  &=  \frac{(\gamma_l^{\bot} )^2 \lambda_l }{
 \bigl[ 1/ {u}_l^{\bot}  + \lambda_l - 
  \gamma_l^{\bot}    \bigr]^2}   - A_l^{\bot}, 
\\
A_l^{\alpha}  &=  (\gamma_l^{\alpha} )^2 \lambda_l   I
 \left( 1/ {u}_l^{\alpha}  + \lambda_l - \gamma_l^{\alpha}   ,
 \gamma_l^{\alpha} -1 ,  \lambda_l \right).
 \label{eq:AlalphanoM}
 \end{align} 
Here, we have introduced the
dimensionless integral
 \begin{align}
 I ( a , b , \lambda )
  &=  \int_0^{\lambda} dx \frac{1}{( a + b x )^2 (1 +x )^2} 
\nonumber 
\\
 &= 
\frac{ \lambda }{  ( a -b )^2 } \left[
 \frac{1}{1 + \lambda } + \frac{ b^2}{a ( a + b \lambda )} \right]
 \nonumber
 \\
   & \quad {} 
- \frac{2 b}{ (a -b)^3}  
 \ln \left[ \frac{ a ( 1 + \lambda )}{ a + b \lambda } \right].
 \label{eq:I2new}
\end{align}
Of special interest is the manifestly spin-rotationally invariant decomposition (\ref{eq:HSrotinv}).
In this case $U^\parallel = U^\bot/2 = U/3$, such that each of the longitudinal and two transverse channels has the same weight. 
In terms of the dimensionless bare couplings 
$u_0^{\parallel}$ and $u_0^{\bot}$
defined in Eq.~(\ref{eq:ulbot})
we therefore have $u_0^{\parallel} = u_0^{\bot} = 2u_0/3$.
It is now easy to see that for any $l$ we have $u_l^{\parallel} = u_l^{\bot} = 2u_l/3$ with $u_l = Z_l u_0$ and $\gamma_l \equiv \gamma_l^\parallel = \gamma_l^\bot$. Our flow equations (\ref{eq:Zlogflow})--(\ref{eq:gammalonlogflow}) then reduce to 
 \begin{align}
 \frac{\partial_l Z_l}{Z_l}   &=  -  \eta_l   ,
 \label{eq:Zlogflowsym}
 \\
 \frac{\partial_l \gamma_l}{\gamma_l }
  &=  \frac12 A_l  ,
 \label{eq:gammalogflowsym}
 \end{align}
where
\begin{align}
 \eta_l  &= \frac{3 \gamma_l^2 \lambda_l  }{2 
 \bigl[ 3/(2 {u}_l) + \lambda_l - 
 \gamma_l \bigr]^{2}}   - \frac{3}{2} A_l ,
 \\
A_l  &=  \gamma_l^2 \lambda_l   I
 \left( 3/(2 {u}_l)  + \lambda_l - \gamma_l   ,
 \gamma_l -1 ,  \lambda_l \right).
 \label{eq:AlalphanoM}
 \end{align} 
The factors of $3$ appearing here are related to the fact that for this choice of parameters we have three equivalent channels.
Clearly, our truncated flow equations respect the spin-rotation symmetry of the problem such that for this special decomposition we can expect the best results.

\subsection{Results}

The three-dimensional system of differential equations given in
Eqs.~(\ref{eq:Zlogflow})--(\ref{eq:gammalonlogflow})
can easily be solved numerically. An example of a typical RG flow 
is shown in Fig.~\ref{fig:RGflow}.
\begin{figure}[tb]    
   \centering
 \epsfig{file=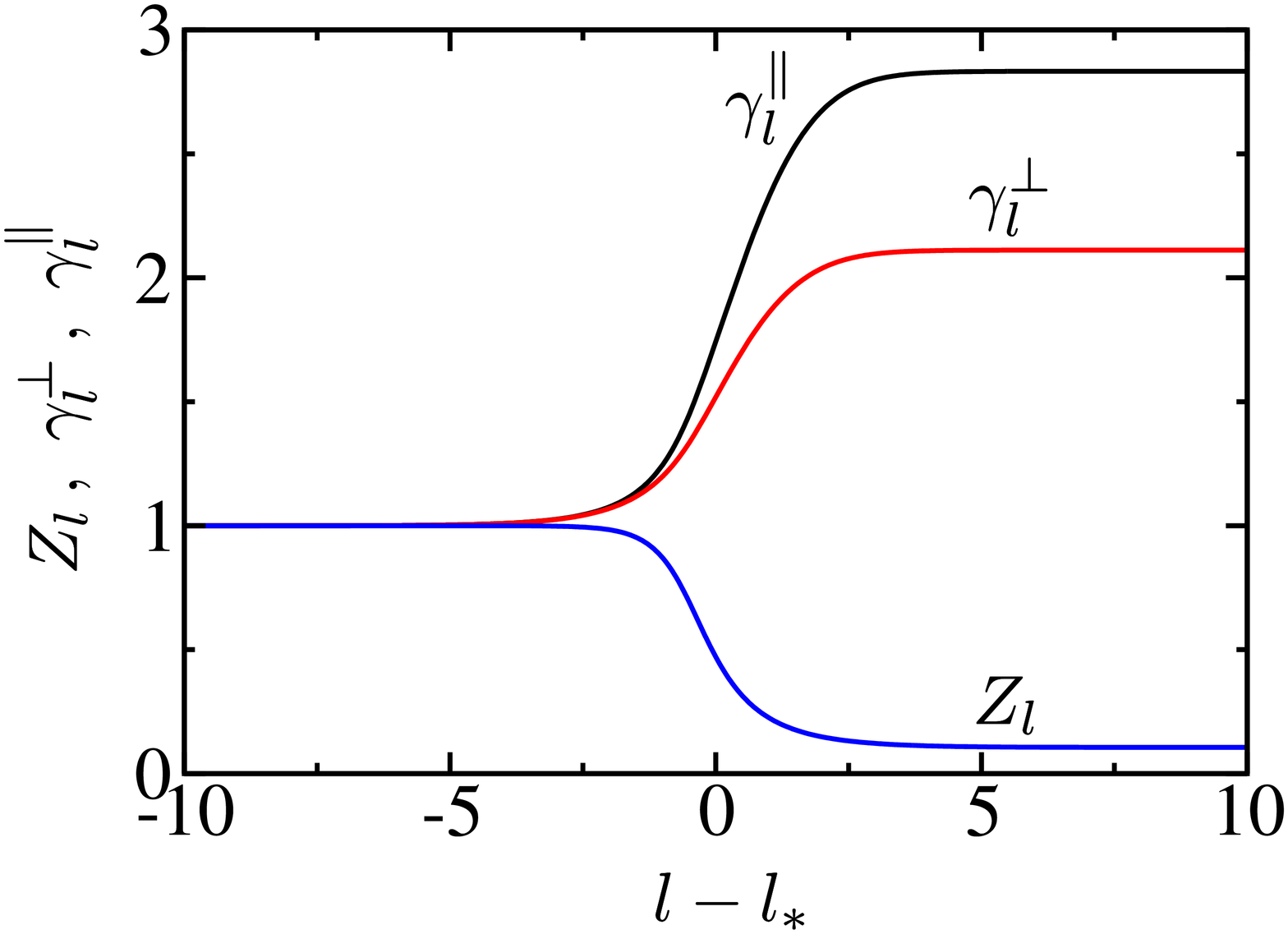,width=70mm}
  \caption{%
(Color online)
Typical flow of the wave function renormalization factor $Z_l$ and of the
frequency-independent parts $\gamma_l^{\bot}$ and $\gamma_l^{\parallel}$
of the boson-fermion vertices obtained from the numerical solution
of Eqs.~(\ref{eq:Zlogflow})--(\ref{eq:gammalonlogflow}).
The scale
$l_{\ast} = \ln ( \Lambda_0 / \Delta )$  is the same as in Fig.~\ref{fig:Zflow}.
The curves are for $u_0^\parallel=1$ and $u_0^{\bot} = 1.5$, such that $u_0 = u_0^\parallel/2 + u_0^\bot =2$.
}
    \label{fig:RGflow}
  \end{figure}
Obviously, the wave function renormalization factor decreases monotonically
as the RG is iterated, while the
vertex corrections $\gamma_l^{\parallel}$ and $\gamma_l^{\bot}$ both increase.
The strongest variations occur at the scale
 $l_{\ast}$ where the effective cutoff
$\Lambda_0 e^{ - l_{\ast}}$ is equal to the hybridization $\Delta$.

Before discussing the behavior of $Z = \lim_{l \rightarrow \infty} Z_l$
in the strong coupling regime, let us fix the optimal choice
of the relative weight of the bare couplings
$U^{\parallel}$ and $U^{\bot}$
in the Hubbard-Stratonovich decoupling (\ref{eq:decomp}).
In Fig.~\ref{fig:HSinitial} we show the dependence of the 
wave function renormalization factor on the choice of
$u_0^{\parallel} / u_0^{\bot}$ for fixed values of $u_0$.
\begin{figure}[tb]    
   \centering
 \epsfig{file=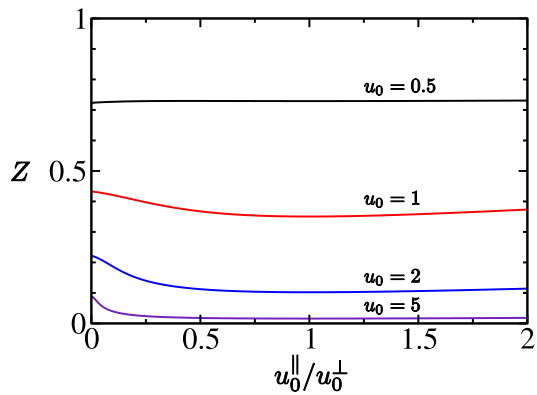,width=70mm}
  \caption{%
(Color online)
Dependence of the quasiparticle residue $Z$ on the choice of
the Hubbard-Stratonovich decoupling, parameterized by
the ratio $u_0^{\parallel} / u_0^\bot$, with $u_0^{\parallel}/2 + u_0^{\bot} = u_0$.
All curves exhibit a local minimum precisely at $u_0^{\parallel} / u_0^\bot = 1$, corresponding to the manifestly spin-rotationally invariant decoupling $u_0^{\parallel} = u_0^\bot = 2 u_0/3$. 
While this minimum is also the global minimum for $u_0 = 1,2$ and $5$, it is only a local minimum for $u_0 = 0.5$. In fact, for $u_0 = 0.5$ the curve is almost flat and has a global minimum at $u_0^{\parallel} / u_0^\bot = 0$ and a local maximum between the two minima.
}
    \label{fig:HSinitial}
  \end{figure}
It turns out that all curves have a pronounced plateau 
with a  minimum precisely at $u_0^{\parallel}/u_0^{\bot} = 1$,
which corresponds
to the manifestly spin-rotationally invariant decoupling (\ref{eq:HSrotinv}).
If we fix the choice of the Hubbard-Stratonovich decoupling by
demanding that first-order variations of the results around 
the optimal choice should vanish, then
we are naturally led to the decoupling (\ref{eq:HSrotinv}).
For the rest of this section we shall therefore use the manifestly spin-rotationally invariant choice $u_0^{\parallel} = u_0^\bot = 2 u_0/3$.

In Fig.~\ref{fig:Z2channel} we show 
our numerical results for $Z$ as a function of the dimensionless bare coupling $u_0$.
\begin{figure}[tb]    
   \centering
 \epsfig{file=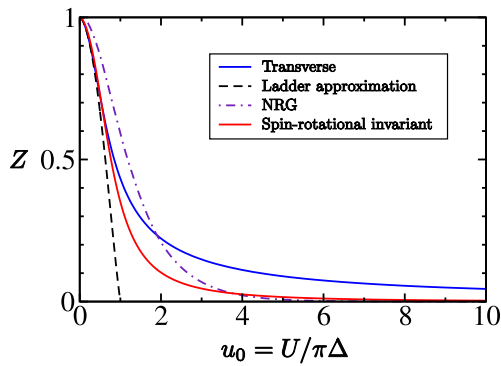,width=70mm}
  \caption{%
(Color online)
Wave function renormalization factor  $Z = \lim_{ l \rightarrow \infty } Z_l$
as a function of the bare coupling $u_0 = U / ( \pi \Delta )$
obtained from the numerical solution of 
Eqs.~(\ref{eq:Zlogflow})--(\ref{eq:gammalonlogflow}).
The two-channel FRG results are for the manifestly spin-rotationally invariant choice
of the two Hubbard-Stratonovich decouplings where $u_0^{\parallel} = u_0^{\bot} = 2 u_0/3$. For a comparison, we have also shown NRG results from Ref.~[\onlinecite{Karrasch08}] (which we have extrapolated to values $u_0 \gtrsim 4.5$.)
}
    \label{fig:Z2channel}
  \end{figure}
For comparison, we also show the prediction of the ladder
approximation, as well as our FRG results with decoupling only in the
transverse spin channel, and numerically accurate results
obtained within Wilson's numerical renormalization group \cite{Karrasch08}.
Our FRG calculation with simultaneous decoupling in both transverse and
longitudinal spin channels obviously yields much better results for the suppression of
$Z$ in the strong coupling regime than the single-channel FRG discussed
in Sec.~\ref{sec:FRG1}. In fact, on the scale of Fig.~\ref{fig:Z2channel}
our FRG results for $Z$ seem  to be reasonably close to the exact numerical results,
which in the strong coupling regime can be approximated by
the asymptotic formula (\ref{eq:Zexact}).
To investigate whether our FRG approach reproduces the
known exponential suppression of $Z$, it is useful to present the data in
Fig.~\ref{fig:Z2channel} by plotting $1/Z$ on a logarithmic scale, as shown in
Fig.~\ref{fig:Zinv2channel}.
\begin{figure}[tb]    
   \centering
 \epsfig{file=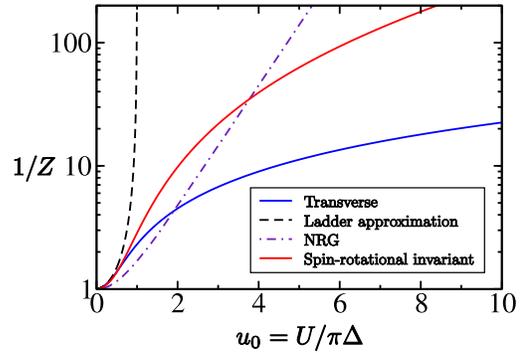,width=70mm}
  \caption{%
(Color online)
Redrawing of Fig.~\ref{fig:Z2channel}:  the
inverse wave function renormalization factor  $1/Z$ is now plotted 
on a logarithmic scale.
}
    \label{fig:Zinv2channel}
  \end{figure}
Note that on this scale the exponential suppression of $Z$ at strong coupling
corresponds to a straight line. Obviously, for $ U \gtrsim 15 \Delta $ our 
two-channel FRG results begin to deviate significantly from the NRG results
and definitely do not reproduce the known exponential suppression of the
quasiparticle weight for $U \rightarrow \infty$. On the other hand,
the two-channel FRG is accurate 
up to  $ U \approx 15 \Delta$. 
We suspect that the deviations from the NRG results for $U \lesssim 15 \Delta$ are due to our linearization of the spin susceptibility, which is expected to lead to a suppression of the quasiparticle weight, see Fig.~\ref{fig:ZUplot}.

Finally, we show in Fig.~\ref{fig:specfunc} the spectral density of the $d$-electrons,
which is defined by
 \begin{equation}
 A ( \omega ) = - \frac{1}{\pi} {\rm Im}\, G ( \omega + i 0 ).
 \label{eq:specdef}
 \end{equation} 
\begin{figure}[tb]    
   \centering
 \epsfig{file=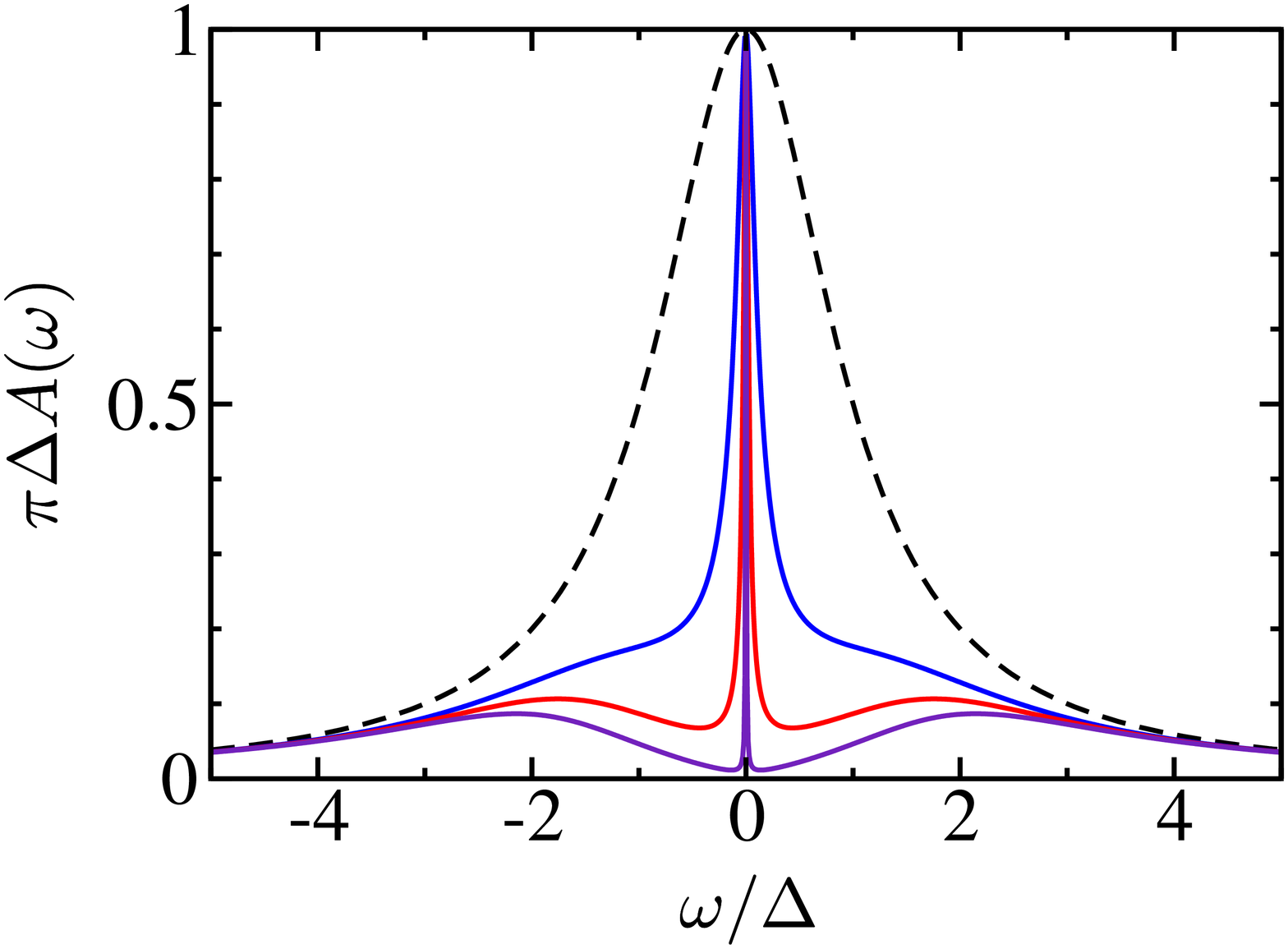,width=70mm}
  \caption{%
(Color online)
Low energy behavior of the spectral density $A ( \omega ) $
of the $d$-electrons as defined in Eqs.~(\ref{eq:specapprox})
for $u_0 = 0$ (dashed line) and $2,4,10$.
}
    \label{fig:specfunc}
  \end{figure}
For simplicity, we approximate $G(\omega+i0)$ by $G_{\Lambda=\omega}(i\omega\to\omega+i0)$ and obtain 
 \begin{equation}
 \pi \Delta A ( \omega ) =  \frac{1}{ 1 + 
\left( \frac{ \omega}{ Z ( \omega ) \Delta }
 \right)^2 },
 \label{eq:specapprox}
 \end{equation}
with $Z ( \omega ) = Z_{ l = \ln ( \Lambda_0 / \omega ) }$.
The qualitative behavior of $A ( \omega )$ 
is in agreement with known results \cite{Hewson93}: 
the width of the central Kondo peak 
is proportional to the wave function renormalization factor $Z$
and its height at zero frequency is pinned to the value $1/( \pi \Delta )$ 
in the particle-hole symmetric case considered here, so that 
the overall spectral weight of the low-energy peak is of order
$Z$. The spectral line-shape at energies larger than $\Delta$ 
(including the broadened Hubbard bands at energy scales $\pm U/2$)
cannot be described within our low-energy truncation.

\section{Summary and conclusions}
\label{sec:summary}

In summary, we have proposed a functional integral approach
to the Anderson impurity model which is based on the mapping
of the original fermionic problem onto a mixed 
Bose-Fermi theory where
the spin fluctuations
are represented by bosonic Hubbard-Stratonovich fields.
Our approach can be used to approximately calculate the
spectral properties including the quasiparticle weight up to couplings $U \lesssim 15 \Delta$. 

This work  also contains several technical advances which will be useful for other FRG 
calculations. In particular, we have shown 
that in FRG calculations for Fermi systems using the technique of
partial bosonization \cite{Schuetz05,Schuetz06,Kopietz09,Baier04,Wetterich07,Strack08},
the skeleton equations in the bosonic sector (which follow from
the Dyson-Schwinger equations) can be used
to close the FRG flow equations in the fermionic sector.
Moreover, we have shown how the ambiguities inherent in multi-field
Hubbard-Stratonovich decouplings can be resolved
if one demands minimal sensitivity with respect to small variations of
a given choice of decoupling. For a symmetric Anderson impurity model,
this naturally leads to the manifestly spin-rotationally symmetric decoupling
(\ref{eq:HSrotinv}) at strong coupling.
With our decoupling scheme, this symmetry is also respected by the truncated flow equations.

 Our approach still has two major shortcomings:\\
(i) We have not been able to reproduce the exponential dependence
$Z \propto \exp[ - \pi^2 u_0 /8]$
of the wave function renormalization factor for $U \rightarrow \infty$.
Although we have tried several modifications of our approach, 
the strong fluctuations responsible for an exponential suppression of $Z$
for $U \rightarrow \infty$ are apparently not correctly
described within our truncation of the exact FRG flow equations.\\
(ii) In order to make progress without using  
complicated numerical  methods, we have made approximations
which are only accurate at low energies $ | \omega | \lesssim \Delta$.
However, a direct numerical solution of our FRG flow equations
should give rise to a much better agreement with NRG data at small couplings $U$ and might also correctly reproduce 
the high-energy behavior of the spectral function.
Therefore one should numerically solve
the coupled system of integro-differential equations
given by Eqs.~(\ref{eq:skeleton2}), (\ref{eq:selfflow22}) or
Eqs.~(\ref{eq:selfflow2})--(\ref{eq:skeletontrapprox}), 
without approximating the fermionic Green function on the
right-hand side of these equations by the
Fermi liquid form (\ref{eq:GFL}).
The fact that with our simple approximations we get results which agree reasonably well with NRG results gives us hope that solving this coupled system of integro-differential equations could lead to very accurate results.

\section*{ACKNOWLEDGMENTS}

We would like to acknowledge useful conversations with W.~Hofstetter, T.~Pruschke and A.~Sinner.
Part of the work by P.~K.  was performed 
during a visit
at the International Center  for Condensed Matter Physics at the
University of Bras\'{i}lia  which was financially 
supported by the DAAD/PROBRAL program.
J. J. R. C.  gratefully acknowledges a  
DAAD postdoc fellowship.

\end{document}